\documentclass[twocolumn,11pt]{aastex62}
\usepackage{amsmath}
\usepackage{graphicx}
\usepackage{CJK}
\usepackage{subfigure}

\long\def\symbolfootnote[#1]#2{\begingroup%
\def\thefootnote{\fnsymbol{footnote}}\footnote[#1]{#2}\endgroup}

\newcommand{\beq}{\begin{equation}}
\newcommand{\eeq}{\end{equation}}
\newcommand{\bea}{\begin{eqnarray}}
\newcommand{\eea}{\end{eqnarray}}

\begin{document}

\title{\Large \bf Estimates of the early EM emission from compact binary mergers}

\begin{CJK*}{UTF8}{gbsn}
\author{Yan Li (李彦)}\thanks{liyan287@mail2.sysu.edu.cn}
\affiliation{School of Physics and Astronomy, Sun Yat-Sen University, Zhuhai, 519082, P. R. China}
\author{Rong-Feng Shen (申荣锋)}\thanks{shenrf3@mail.sysu.edu.cn}
\affiliation{School of Physics and Astronomy, Sun Yat-Sen University, Zhuhai, 519082, P. R. China}

%%%%%%%%%%%%%%%%%%%%%%%%%%%%%%%%%%%%%%%%%%%%%%%%%%%%%%%%%%%%%%%%%%%%%%%%%
\begin{abstract}

Compact binary mergers that involve at least one neutron star, either binary neutron star or black hole--neutron star coalescences, are thought to be the potential sources of electromagnetic emission due to the material ejected during the merger or those left outside the central object after the merger. Since the intensity of these electromagnetic transients decay rapidly with time, one should pay more attention to early emissions from such events, which are useful in revealing the nature of these mergers. In this work, we study the early emission of kilonovae, short $\gamma$-ray bursts and cocoons that could be produced in those mergers. We estimate their luminosities and time scales as functions of the chirp mass which is the most readily constrained parameter from the gravitational wave detections of these events. We focus on the range of chirp mass as $1.3M_{\odot} -2.7M_{\odot}$ which is compatible with one of the merging component being a so-called `mass gap' black hole. We show that the electromagnetic observation of these transients could be used to distinguish the types of the mergers when the detected chirp mass falls in the range of $1.5M_{\odot}-1.7M_{\odot}$. Applying our analysis to the sub-threshold GRB GBM-190816, we found that for this particular event the effective spin should be larger than 0.6 and the mass of the heavier object might be larger than 5.5$M_{\odot}$ for the SFHo equation of state.

\end{abstract} 

%%%%%%%%%%%%%%%%%%%%%%%%%%%%%%%%%%%%%%%%%%%%%%%%%%%%%%%%%%%%%%%%%%%%%%%%%%%
\section{Introduction}

The first discovery of a binary neutron star (BNS) merger in GWs made by Advanced LIGO $\&$ VIRGO, GW170817, opened the muti-message astronomy era \citep{abbott17a}. Besides the information encoded in the gravitational wave (GW) strain data alone, the observation of an electromagnetic (EM) counterpart in coincidence with the GW chirp could reveal a much richer picture of these events \citep{bloom09}. 

A compact binary merger that involve at least one neutron star (NS), either BNS or black hole (BH)--NS (BH--NS), would produce a non-negligible fraction of ejecta when the NS is torn apart by the BH's tidal forces or collides with another NS. There are mainly two components of neutron-rich matter that can be ejected. One is the dynamically ejected matter on the dynamical timescales (typically milliseconds) during the merger, which is referred as dynamical ejecta. The other is unbound from the merger remnant disk by neutrino-driven and/or viscously-driven winds, which is referred as disk wind. Those neutron-rich ejecta would produce various kinds of EM emission, as detailed below.

The unbound neutron-rich ejecta from a BNS or BH--NS merger undergoes rapid neutron capture (r-process) nucleosynthesis which results in the formation of heavy elements like gold and platinum. The radioactive-decay heating of the ejecta by those unstable isotopes produces so-called kilonova emission \citep{li98}. The detailed emission properties depend on the ejecta mass, velocity, and composition. Heavier elements known as lanthanides (formed in low electron-fraction material) can increase the ejecta opacity by several orders of magnitude \citep{kasen13,tanaka13}, making the kilonova light curve fainter, redder, and longer-lived \citep{barnes13, grossman14}. The greater the mass of material that is ejected, the brighter the transient that becomes, and the longer the timescale on which it will peak.

BNS and BH--NS mergers with appropriate parameters (e.g., BH spin, mass ratio, equation of state (EOS), etc.) are likely accompanied by short $\gamma$-ray bursts (sGRBs) \citep{eichler89, narayan92, fong14}. The first BNS merger event GW170817 was followed by a sGRB, i.e., GRB170817A \citep{abbott17b}. \cite{shapiro17}, \cite{paschalidis17} and \cite{ruiz18} have performed simulations of merging systems in full general relativistic magnetohydrodynamics (MHD) and showed that after a BH--NS merger a relativistic jet can be launched, powering a sGRB. In the BH--NS merger scenario, the central engine of the accompanying sGRB is the accretion of disk onto the spinning black hole in the Blandford-Znajeck (BZ) mechanism \citep{blandford77}. Thus the luminosity and the duration of the burst depend on the properties of the disk.

A hot, mildly relativistic cocoon surrounding the jet could be generated by the interaction of the lighter sGRB jet with the denser material ejected previously (i.e., magnetic/viscous/neutrino-driven wind from the disk and/or dynamical ejecta) along the rotation axis of the remnant BH formed after a BNS or BH--NS merger \citep{murguia14, murguia17, nagakura14}. The energy deposited by the shock in the cocoon diffuses as it expands and escapes to the observer, producing a cooling X-ray, UV and optical emission \citep{nakar17, lazzati17}. The prompt emission of the cocoon is detectable for Swift BAT and Fermi GBM \citep{lazzati17}.

The intensity and timescales of these early EM transients strongly depend on the amount of matter that is ejected during merger, bound in an accretion disk, or ejected in the form of post-merger disk winds, all of which depend on the properties of the coalescing compact objects \citep{foucart12, kawaguchi16, dietrich17, radice18}. \cite{foucart18} introduced a fitting formula that estimates the remnant mass (including disk as well as dynamical ejecta) for BH--NS mergers. \cite{kruger20} provided updated fitting formulae that estimate the disk mass for BNS and dynamical ejecta masses for BH--NS and BNS, fitted to the results of numerical simulations. Combining their results, one can make a rough estimate of the two ejecta components in different merger events, and hence give some estimates for those early EM emission, which can be used (in conjunction with other observations) to constrain the binary system parameters of an observed merger event.

Due to the large distance and the rapid decaying nature of those EM emissions, one should pay more attention to early (several hrs) emissions from such events, which are brighter than that afterward. Therefore, we study the properties of the early EM emission (e.g., kilonova, sGRB, cocoon) from compact binary mergers and derive some estimates of them.

As for a detected GW source, the binary chirp mass $M_{ch}$, which combines the masses of the two components (see Eq. 5 for its definition), is one of the best parameters constrained from the gravitational signal in low-latency searches. It is the primary parameter used to reveal the nature of the binary, such as whether the system hosts two NSs, two stellar BHs or a BH and a NS. \cite{barbieri19a} pointed out that for some range of the chirp mass value, using chirp mass alone cannot distinguish a BNS merger from a BH--NS merger. In this case, for some selected chirp masses, the observation of the kilonova emission from these systems can help identify the nature of the merger system \citep{barbieri19a}.

\cite{barbieri19a} studied kilonova emissions from mergers of binary whose chirp mass in the range of $1.2M_{\odot}-1.8M_{\odot}$. In this paper, we investigate the differences in the kilonova emissions between the BNS and BH--NS mergers at given chirp masses which is similar to the approach of \cite{barbieri19a}. However, our studied BNS and BH-NS mergers have a wider range of chirp masses, i.e., $1.3M_{\odot}-2.7M_{\odot}$. Besides kilonova emission, we investigate the properties of the sGRB and cocoon for the BNS and BH--NS mergers and estimate their EM emissions as a function of the chirp mass. For the range of the studied chirp masses, some of the BHs fall in the socalled `mass gap' \citep{bailyn98,ozel10,farr11}, which is defined empirically as $3M_{\odot}-5M_{\odot}$ for the absence of low mass black holes in the Galaxy. \cite{thompson19} recently reported the discovery of a BH with mass $3.3_{-0.7}^{+2.8}M_{\odot}$ in a noninteracting binary system with a red giant. Due to the presence of `mass gap' BH, this work is useful to constrain the properties of the BH--NS merger systems and identify whether an observed BH--NS merger event involves a `mass gap' BH or not.

This paper is structured as follows. In Section 2 we describe the properties of the two components of ejecta for both BNS and BH--NS mergers in more detail. Our general method to derive the masses of dynamical ejecta and disk for the given chirp masses is described in Section 3. Our calculations for the luminosities and timescales of kilonova, sGRB, and cocoon are presented and analyzed in Section 4, 5, and 6, respectively. A case study for GRB GBM--190816 is presented in Section 7. In Section 8 we summarize the results and discuss the implications. 

%%%%%%%%%%%%%%%%%%%%%%%%%%%%%%%%%%%%%%%%%%%%%%%%%%%%%%%%%%%%%%%%%%%%%%%%%%%%
\section{Ejecta properties}

The EM emissions are determined by the properties of the ejecta produced by the BNS or BH--NS mergers. Specifically, the greater the mass of the ejecta is, the brighter the EM transient and the longer the peak timescale will be. Besides, a higher opacity of the ejecta would delay and redden the kilonova transient. For BNS and BH--NS mergers, there are mainly two components of neutron-rich ejecta, i.e., dynamical ejecta and disk wind, whose properties are described in the following subsections.

%%%%%%%%%%%%%%%%%%%%%%%%%%%%%%%%%%%%%%%%%%%%%
\subsection{Dynamical ejecta}

In a BNS merger, the tidal force peels matter from the surface of the approaching star, producing outward expanding tails of debris which are primarily in the equatorial plane \citep{rosswog99}. When the stars come into contact, shock heating at the interface could squeeze additional matter  into polar regions \citep{oechslin07, sekiguchi16}. The ejecta properties depend sensitively on the fate of the massive NS remnant, the total binary mass, the mass ratio of the binary, and the EOS (see \cite{fernandez16, shibata19}, for recent reviews). For BNS mergers, the total masses and velocities of dynamical ejecta typically lie in the range $10^{-4}M_{\odot}-10^{-2}M_{\odot}$ \citep{hotokezaka13a, radice16, bovard17} and $0.1c-0.3c$, respectively.

In a BH-NS merger, when the NS is disrupted prior to the merger, there is substantial matter ejected by the tidal force of the BH. The dynamical ejecta emerges primarily in the equatorial plane and shows a significant anisotropy, different from a BNS merger whose dynamical ejecta is nearly isotropic \citep{Kawaguchi15, just15}. \cite{foucart14}, \cite{kyutoku15} and \cite{kyutoku18} performed numerical-relativity simulations of BH-NS mergers, and found that the dynamical ejecta is concentrated in the equatorial plane with a half opening angle of $10^{\circ}-20^{\circ}$ and sweeps out about a half of the plane. The properties of dynamical ejecta for BH--NS mergers mainly depend on the BH spin, the NS EOS and the mass ratio \citep{fernandez15a,fernandez16,shibata19}. The ejecta mass ranges from $0.01M_{\odot}$ to $0.1M_{\odot}$, and the average velocity of the ejecta inferred from the kinetic energy is typically $0.2c-0.3c$. This ejecta is highly neutron rich, producing heavy r-process elements (e.g., Lanthanides) which result in a high opacity $\kappa\sim$10 $\rm{cm^2/g}$ \citep{kasen17}.

%%%%%%%%%%%%%%%%%%%%%%%%%%%%%%%%%%%%%%%%%%%%%
\subsection{Disk wind}

For all the BNS mergers and some of the BH--NS mergers, the debris of the tidal-disrupted NS(s), which is bound to the central NS or BH remnant, will circularize into an accretion disk. The disk mass with typical value of $\sim 0.01M_{\odot}-0.3M_{\odot}$ \citep{oechslin06, hotokezaka13b, foucart14} depends  primarily on the total mass and mass ratio of the binary, the spins of the binary components, and the NS EOS \citep{oechslin07,shibata11}. After the merger, those neutron-rich matter hung up in an accretion disk around the central merged remnant is partially blown away in the form of winds through neutrino heating and/or viscously driven process \citep{metzger17}. Referred as disk wind, it is another important source of ejecta mass, and often dominate that of the dynamical ejecta. \cite{kiuchi15} performed a high resolution magnetohydrodynamic simulation for BH--NS merger and found that the fraction of the disk mass that can be ejected is up to 50\%. Notably, the amount of the disk outflows varies with the spin of the BH remnant \citep{fernandez15b}. 

As the relatively-slow disk outflows emerge after the dynamical ejecta, they are physically located behind the dynamical ejecta, and possess a more isotropic geometry \citep{metzger17}. The disk wind is generally less neutron-rich because of longer exposure to weak interactions \citep{fernandez17}. \cite{just15} found that the majority of the disk wind material ends up in producing $A <130$ r-process elements, and the disk outflow complements the dynamical ejecta by contributing the lighter r-process nuclei $(A <140)$ that are underproduced by the latter. As a result, the opacity of the disk wind should be lower than that of the dynamical ejecta.

%%%%%%%%%%%%%%%%%%%%%%%%%%%%%%%%%%%%%%%%%%%%%

%%%%%%%%%%%%%%%%%%%%%%%%%%%%%%%%%%%%%%%%%%%%%%%%%%%%%%figure_1

\begin{figure*}[htb]
  \centering
    \subfigure{
    \begin{minipage}[t]{8.6cm}
    \includegraphics[width=8.6cm]{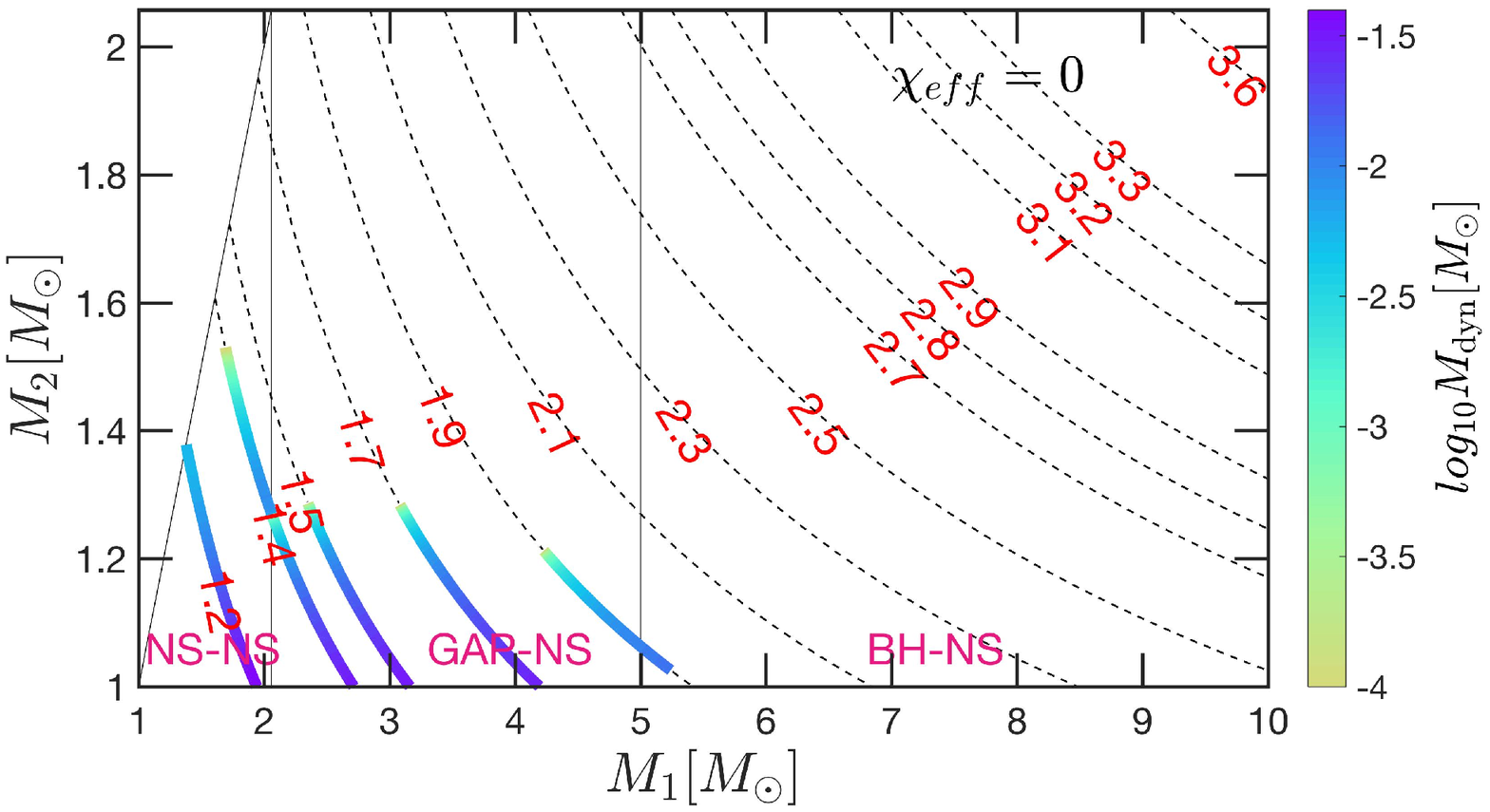}
    \end{minipage}} 
    \subfigure{
    \begin{minipage}[t]{8.6cm}
    \includegraphics[width=8.6cm]{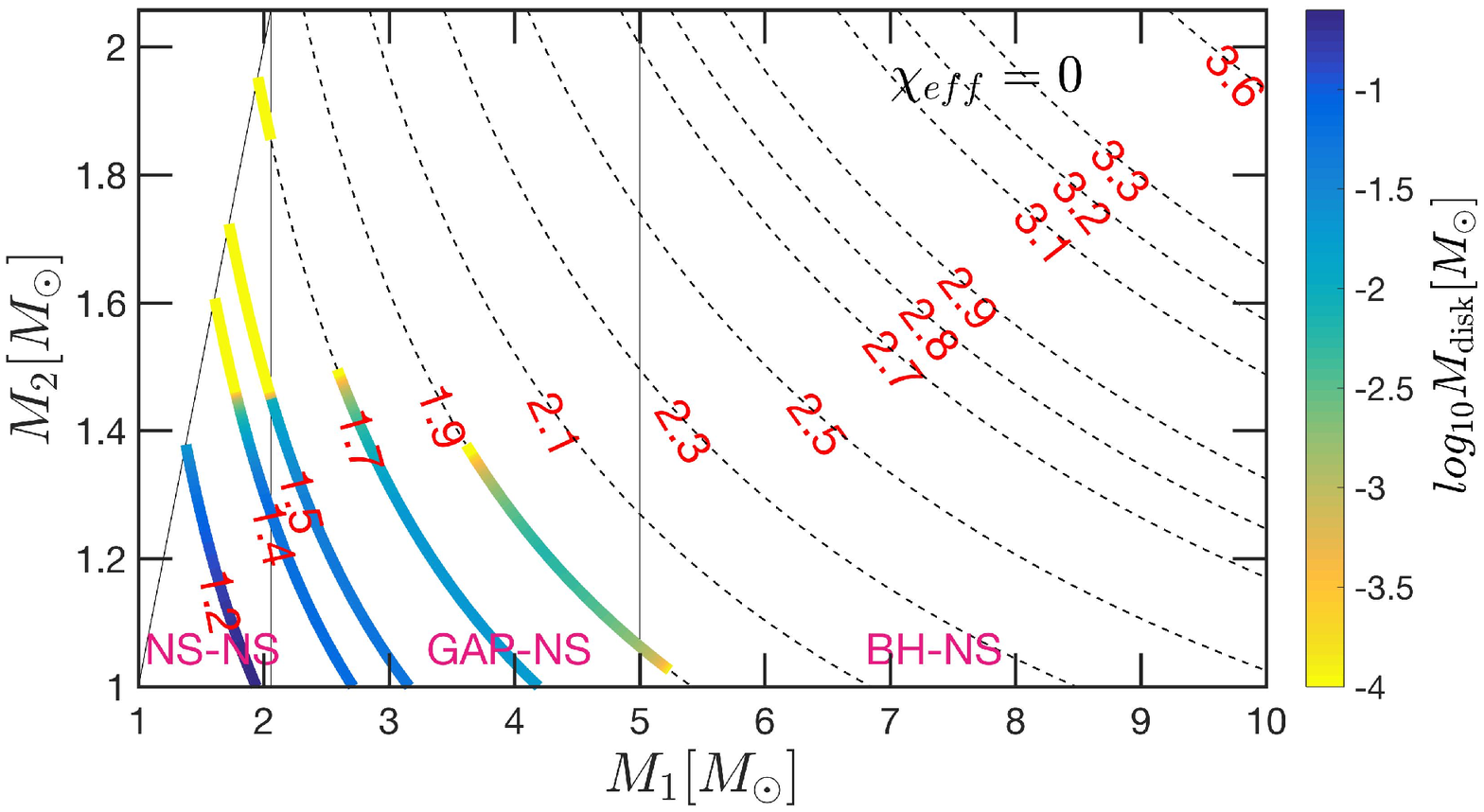}
    \end{minipage}} 
    
    \subfigure{
    \begin{minipage}[t]{8.6cm}
    \includegraphics[width=8.6cm]{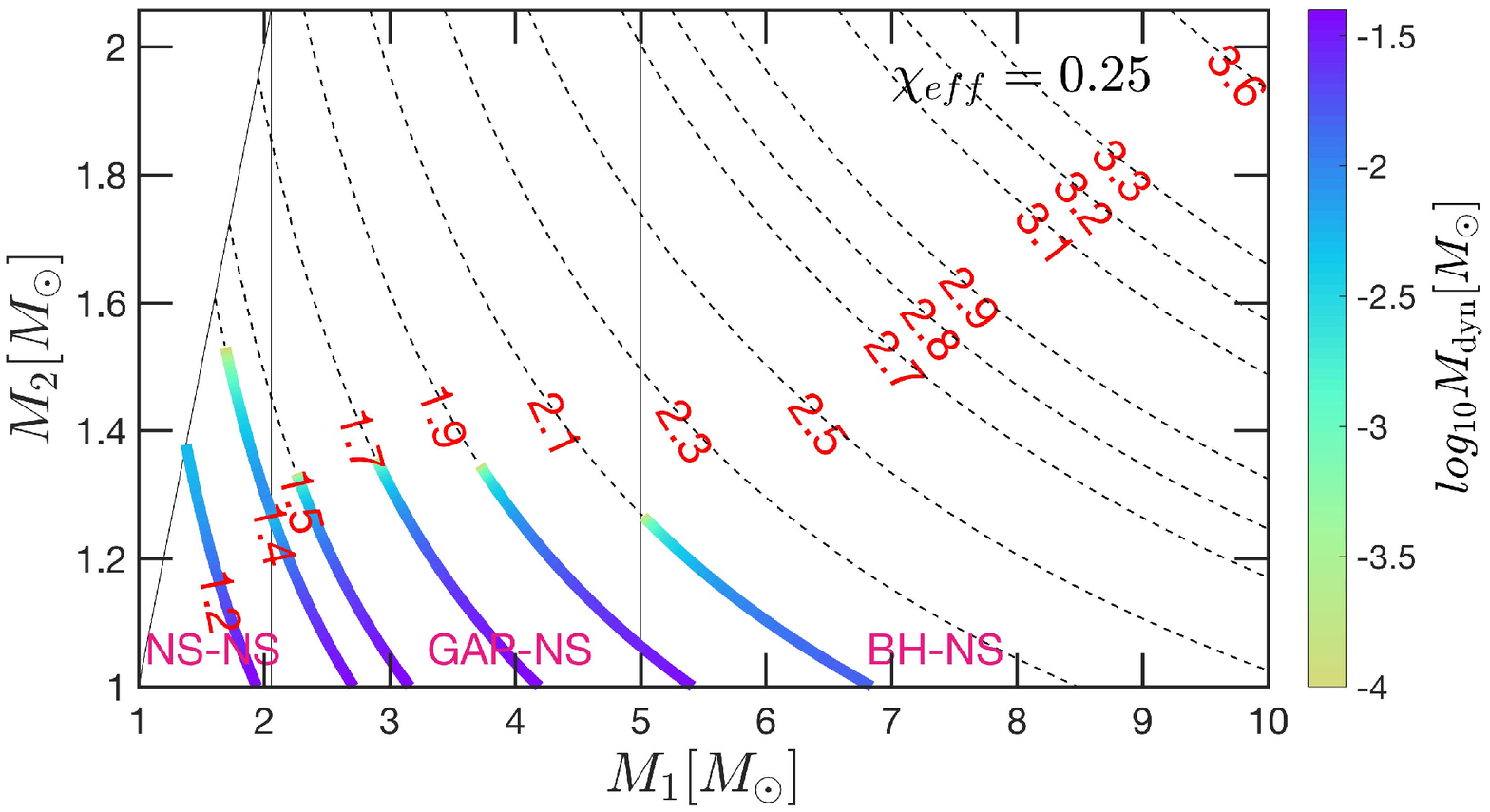}
    \end{minipage}} 
    \subfigure{
    \begin{minipage}[t]{8.6cm}
    \includegraphics[width=8.6cm]{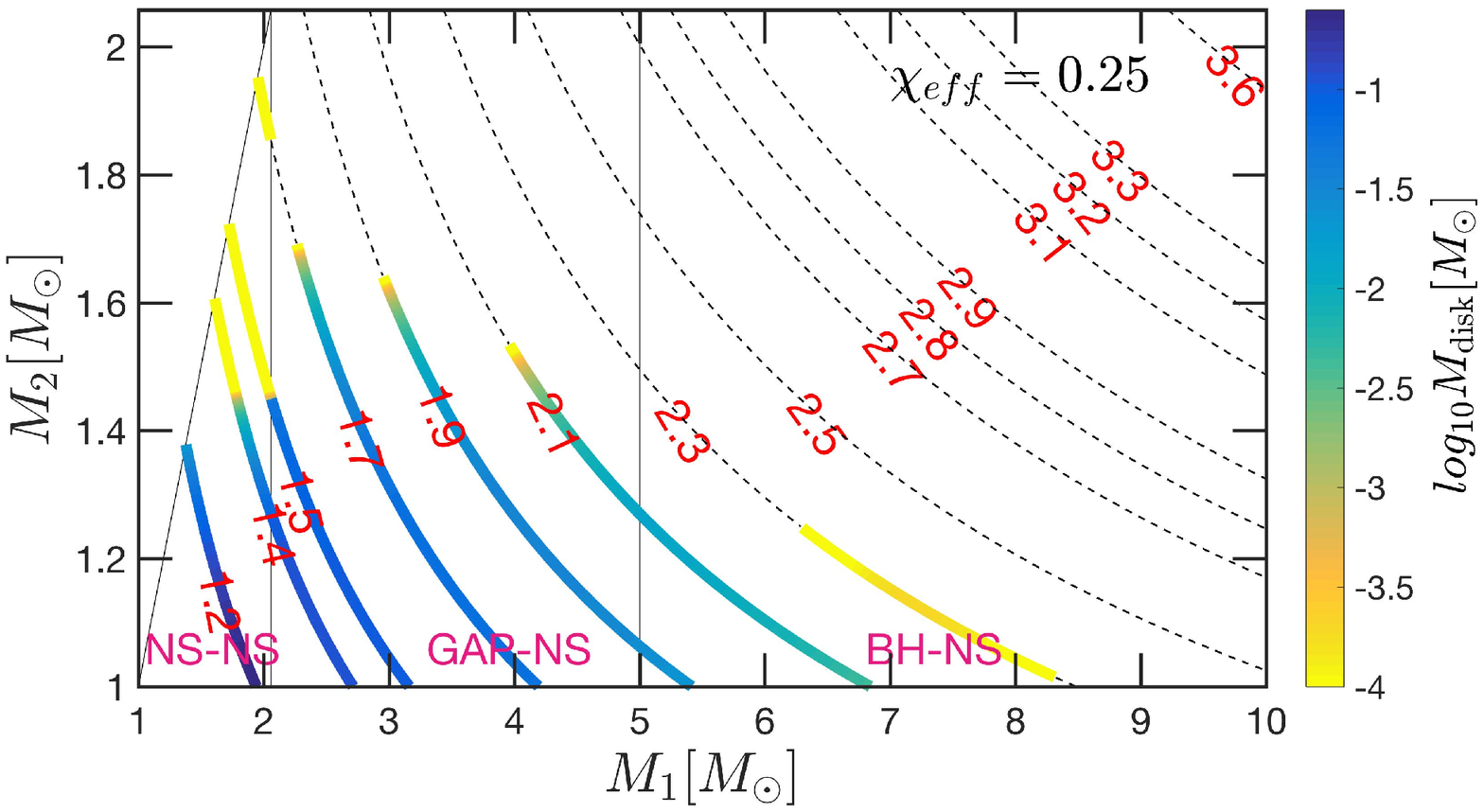}
    \end{minipage}}    
    	
    \subfigure{
    \begin{minipage}[t]{8.6cm}
    \includegraphics[width=8.6cm]{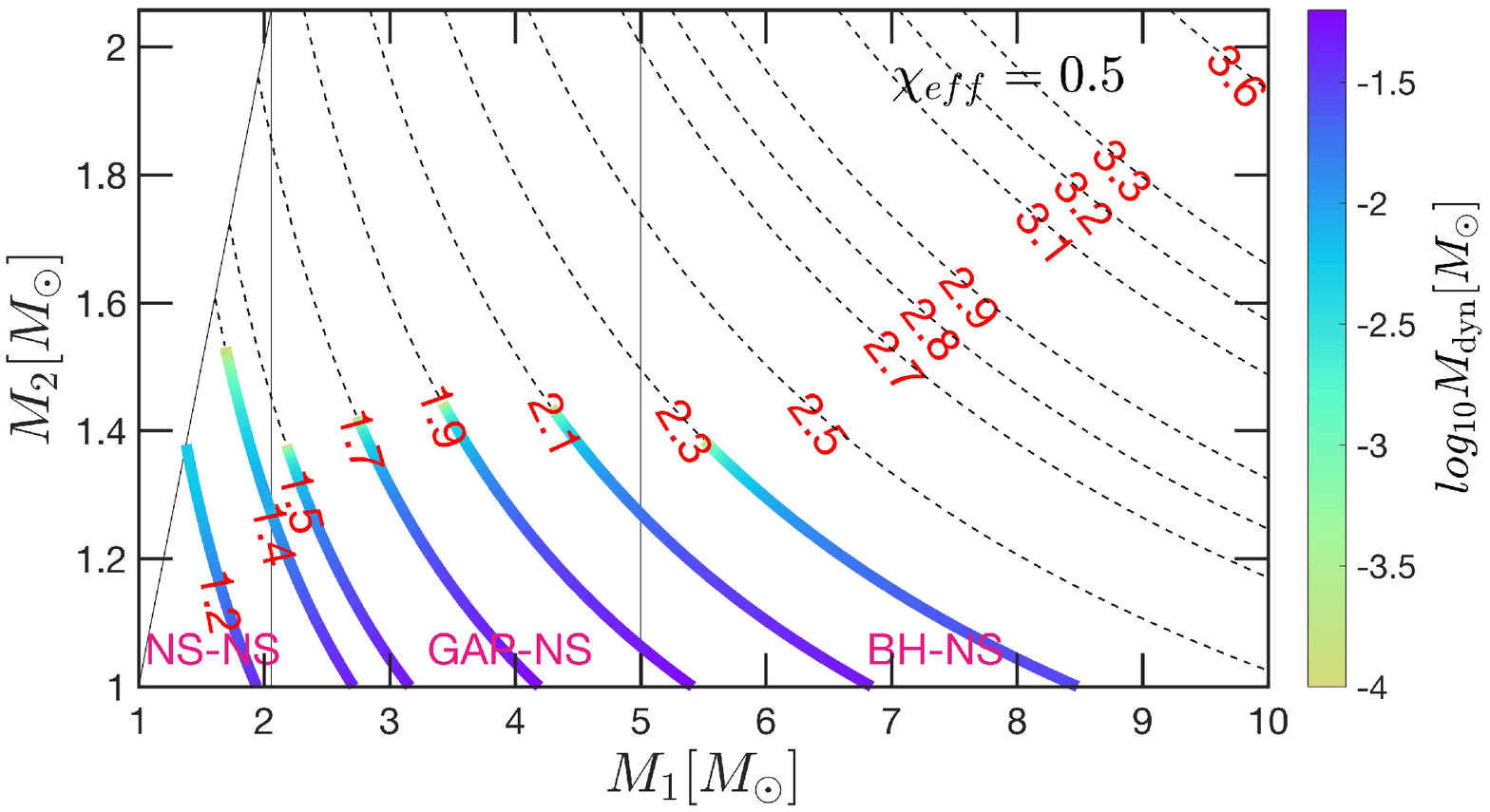}
    \end{minipage}} 
    \subfigure{
    \begin{minipage}[t]{8.6cm}
    \includegraphics[width=8.6cm]{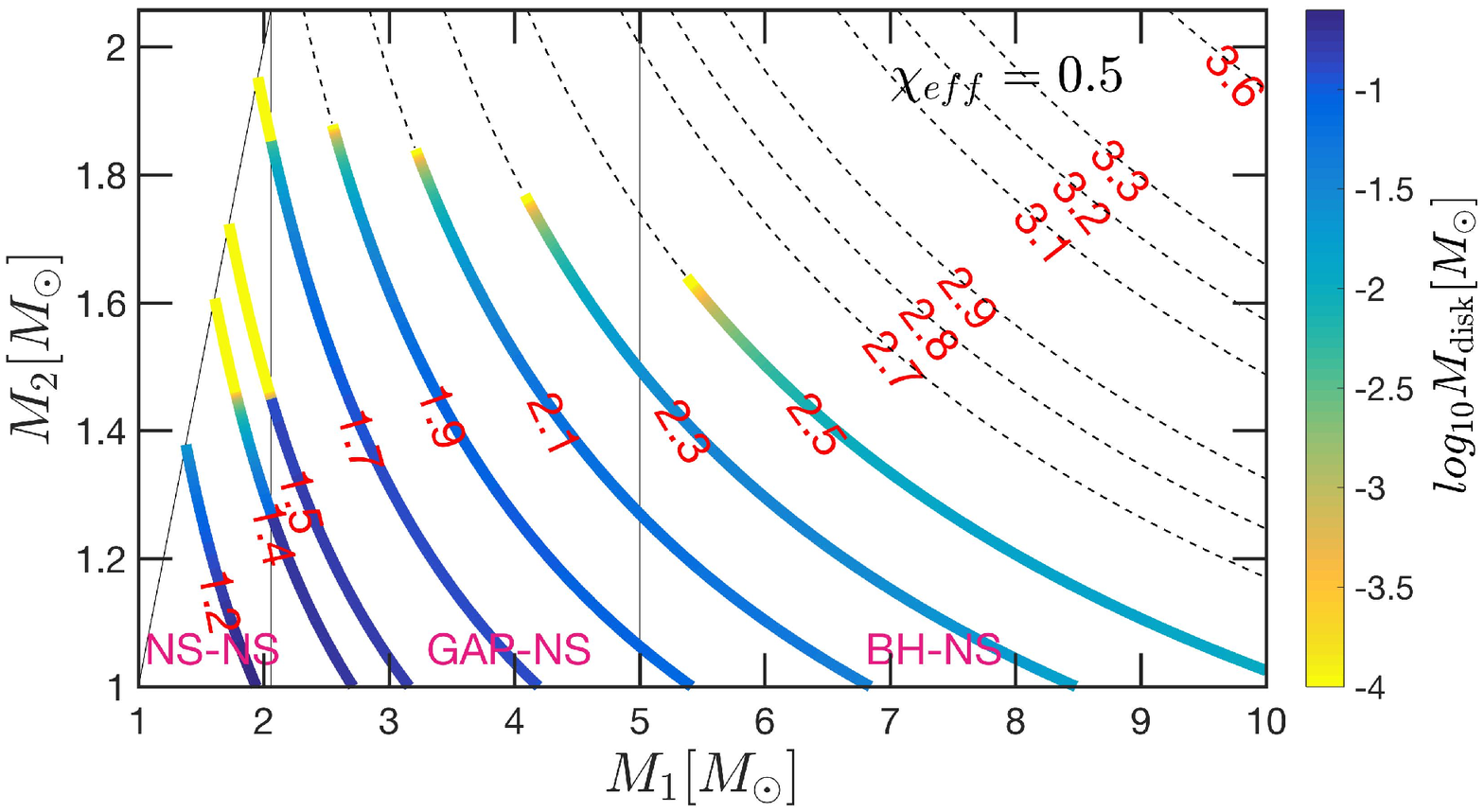}
    \end{minipage}} 
    
    \subfigure{
    \begin{minipage}[t]{8.6cm}
    \includegraphics[width=8.6cm]{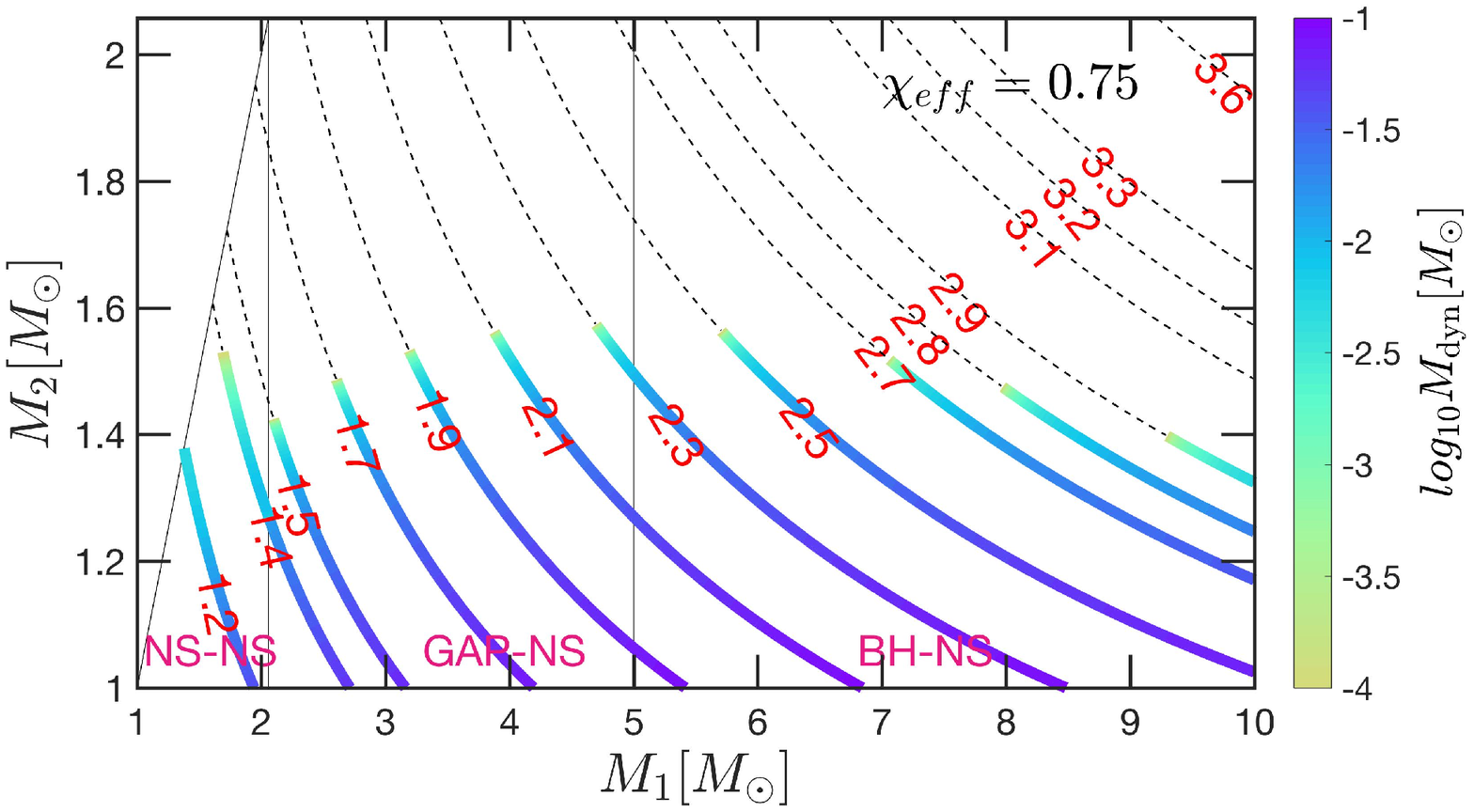}
    \end{minipage}} 
    \subfigure{
     \begin{minipage}[t]{8.6cm}
    \includegraphics[width=8.6cm]{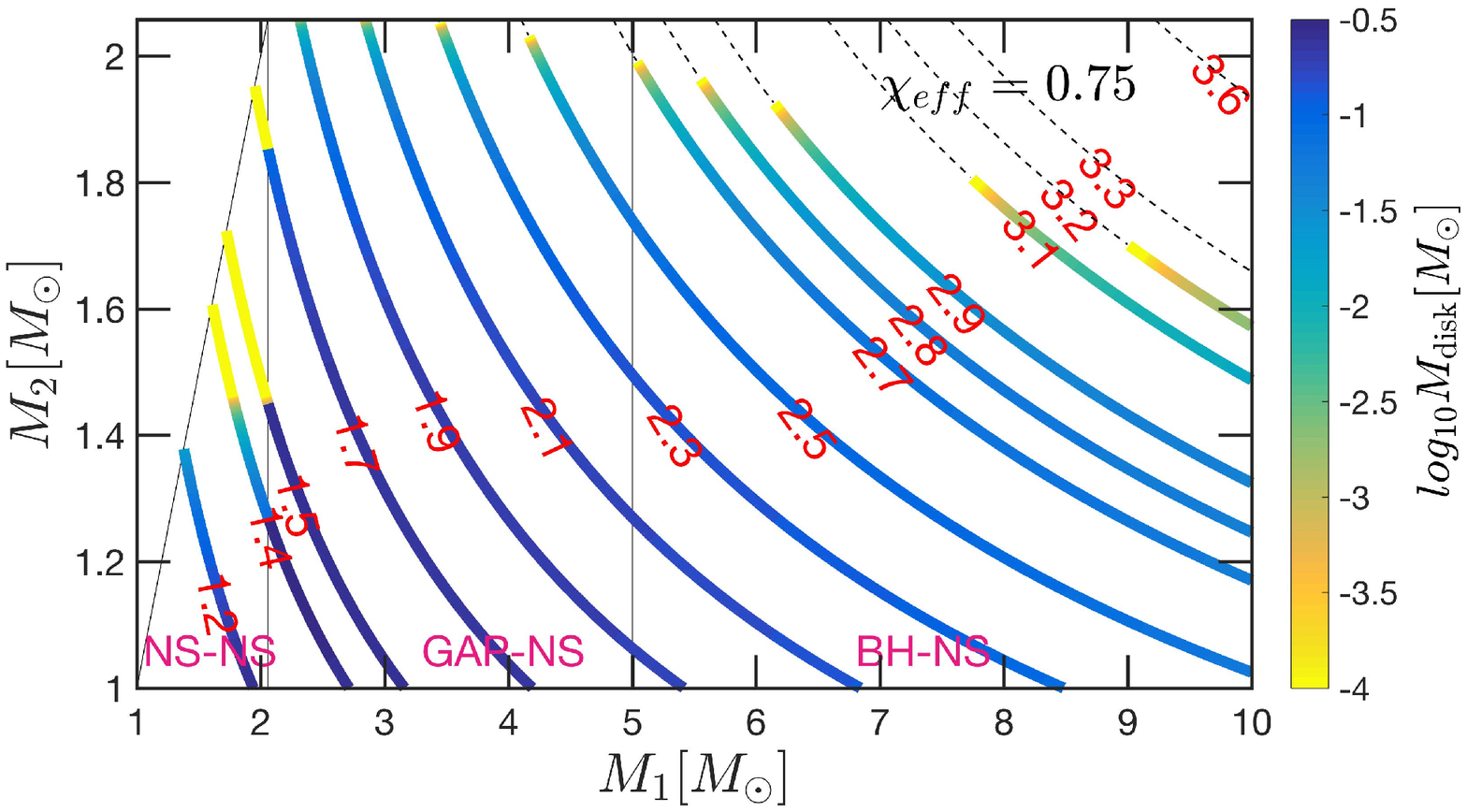}
    \end{minipage}}    
    \centering 
   \caption{Dynamical ejecta (left) and disk (right) masses for given chirp mass $M_{\rm ch}$ whose values are shown in red numbers. From top to bottom, rows of panels correspond to effective BH spin of 0, 0.25, 0.5 and 0.75,  respectively. The dashed lines represent the parameter space where there is no dynamical ejecta or disk.}
	\label{fig:mass_result}
\end{figure*}
%%%%%%%%%%%%%%%%%%%%%%%%%%%%%%%%%%%%%%%%%%%%%%%%%%%%%figure_1

%%%%%%%%%%%%%%%%%%%%%%%%%%%%%%%%%%%%%%%%%%%%%%%%%%%%%%%%%%%%%%%%%%%%%%%%%%%%
\section{method}

\subsection{Estimations of the ejecta mass}

In order to compute the masses of dynamical ejecta and disk from BNS mergers and BH--NS mergers, we adopt the fitting formulae reported in \cite{kruger20} and \cite{foucart18}.  

%%%%%%%%%%BNS dy ejecta
\subsubsection{Dynamical ejecta for BNS mergers}
The dynamical ejecta for BNS mergers can be estimated as \citep{kruger20}
\begin{equation}
\frac{{{M_{{\rm{dyn}}}}}}{{{{10}^{ - 3}}{M_ \odot }}} = \left( {\frac{a}{{{C_2}}} + b\frac{{M_1^n}}{{M_2^n}} + c{C_2}} \right){M_2} + \left( {1 \leftrightarrow 2} \right)
\end{equation}
where $C_2=GM_2/(R_2c^2)$ is the compactness of the lighter of the two NSs, the best-fit coefficients are $a = -9.3335$, $b = 114.17$, $c = -337.56$ and $n = 1.5465$.
%%%%%%%%%%BNS dy ejecta

%%%%%%%%%%BNS disk
\subsubsection{Disk for BNS mergers}
A rather simple fitting formula allows us to predict the disk mass from BNS mergers to good accuracy \citep{kruger20}:
\begin{equation}
{M_{{\rm{disk}}}} = {M_2} \times {\left[\max \left( {a{C_2} + c,5 \times {{10}^{ - 4}}} \right)\right]}^d
\end{equation}
where $a = -8.1324$, $c = 1.4820$ and $d = 1.7784$. This equation is reliable when the mass ratio $q$ ranges from 0.775 to 1. 
%%%%%%%%%%BNS disk

%%%%%%%%%%%%%%%%BHNS ejecta
\subsubsection{Dynamical ejecta for BH--NS mergers}

\cite{kruger20} gave a fitting formula for the dynamical ejecta from BH--NS merger as 
\begin{equation}
\frac{{{M_{{\rm{dyn}}}}}}{{M_{{\rm{NS}}}^b}} = {a_1}{Q^{{n_1}}}\frac{{1 - 2{C_{{\rm{NS}}}}}}{{{C_{{\rm{NS}}}}}} - {a_2}{Q^{{n_2}}}{{\hat{R}_{{\rm{ISCO}}}}} + {a_4}
\end{equation}
where $M_{NS}^b$ and $C_{NS}$ is the baryonic mass and compactness of NS, $Q=M_{BH}$/$M_{NS}$, ${{\hat R}_{{{ISCO}}}} = 3 + {Z_2} - {\mathop{ sign}} \left( {{\chi _{{\rm{eff}}}}} \right)\times\sqrt {\left( {3 - {Z_1}} \right)\left( {3 + {Z_1} + 2{Z_2}} \right)}$, ${Z_1} = 1 + {\left( {1 - \chi _{{\rm{eff}}}^2} \right)^{{1 \mathord{\left/
 {\vphantom {1 3}} \right.
 \kern-\nulldelimiterspace} 3}}}\left[ {{{\left( {1 + {\chi _{{\rm{eff}}}}} \right)}^{{1 \mathord{\left/
 {\vphantom {1 3}} \right.
 \kern-\nulldelimiterspace} 3}}} + {{\left( {1 - {\chi _{{\rm{eff}}}}} \right)}^{{1 \mathord{\left/
 {\vphantom {1 3}} \right.
 \kern-\nulldelimiterspace} 3}}}} \right]$, ${Z_2} = \sqrt {3\chi _{{\rm{eff}}}^2 + Z_1^2}$, $a_1$ = 0.007116, $a_2$ = 0.001436, $a_4$ = $-$0.02762, $n_1$ = 0.8636, $n_2$ = 1.6840, and $\chi_{eff}=\chi_{BH} cos(i_{tilt})$, and $i_{tilt}$ is the angle between the BH spin and the orbital angular momentum.
 
The value of $\chi_{eff}$ is dependent on the formation channel of the binary system. For binaries formed from isolated binary evolution channel, the spin of each component is almost aligned with the orbital angular momenta of the binaries, although modest misalignment may be caused by adequately strong supernova kicks \citep{oshaughnessy17, gerosa18, bavera19}. On the other hand, for binaries produced dynamically in dense stellar environments, their spins have random orientation \citep{kalogera00,mandel10,doctor19}. \cite{abbott20} investigate the population properties of the 47 compact binary mergers detected in GWTC-2, of which 44 are binary black hole (BBH) events. They show that the possibility distribution of $i_{tilt}$ of BH component spins with respect to their orbital angular momenta for BBH systems peaks at $i_{tilt}=0$ (see figure 10 in their paper). Therefore, in this work we adopt $i_{tilt}=0$ for simplicity, and hence the value of $\rm \chi_{eff}$ can be interpreted as the spin of BH.
%%%%%%%%%%%%%%%%%BHNS ejecta

%%%%%%%%%%BHNS disk
\subsubsection{Disk for BH--NS mergers}
\cite{foucart18} introduced a fitting formula for the remnant baryon mass in BH--NS mergers:
\begin{equation}
\frac{M_{\rm rem}}{M_{\rm NS}^b} = {\left[ {\max \left( {\alpha \frac{{1 - 2{C_{{\rm{NS}}}}}}{{{\eta ^{{1 \mathord{\left/
 {\vphantom {1 3}} \right.
 \kern-\nulldelimiterspace} 3}}}}} - \beta {{\hat R}_{{\rm{ISCO}}}}\frac{{{C_{{\rm{NS}}}}}}{\eta } + \gamma ,0} \right)} \right]^\delta }
\end{equation}
where $\eta$ = Q/(1+$Q^2$), $\alpha$ = 0.406, $\beta$ = 0.139, $\gamma$ = 0.255, $\delta$ = 1.761.

Combining Eqs. (3) and (4), as in $M_{disk}=M_{rem}-M_{dyn}$, one can get the mass of disk for BH--NS merger.
%%%%%%%%%%BHNS disk

%%%%%%%%%%%%%%%%%%%%%%%%%%%%%%%%%%%%%%%%%%%%%%%%%%%%%%%%%%%%%%%%%%%%%%%%%%%%
\subsection{Ejecta mass for a given chirp mass}

The binary chirp mass is defined as 
\begin{equation}
{M_{\rm ch}} = \frac{{{{\left( {{M_1}{M_2}} \right)}^{3/5}}}}{{{{\left( {{M_1} + {M_2}} \right)}^{1/5}}}}
\end{equation}
where $M_1$ and $M_2$ are the masses of the two component stars. Here we take $M_1$ $>$ $M_2$. 

LIGO Scientific collaboration \& Virgo Collaboration (LVC) classify merging systems as follows:  `BNS' when both $M_1$ and $M_2$ are less than $3M_{\odot}$, `BBH' when both $M_1$ and $M_2$ are larger than $5M_{\odot}$,  `BHNS' when $M_1>5M_{\odot}$ and $M_2<3M_{\odot}$, or `MassGap' when at least one component possesses a mass between $3M_{\odot}$ and $5M_{\odot}$. 

NS EOS plays an important role on merger process and hence on the mass ejection and disk formation. In this paper, we adopt a soft EOS, i.e., SFHo EOS \citep{steiner13}, whose maximum mass of neutron star is $M_{{\rm{NS}}}^{\max }=2.058 M_{\odot}$. The NS with stiffer EOS possesses a more uniform density profile and will be more susceptible to mass shedding, tidal deformation and tidal disruption by the BH tidal field \citep{shibata11}. Thus our calculation actually gives a conservative estimate for the mass ejection compared to using other stiffer EOS. Note that for the use of the SFHo EOS, we have a new classification different from that of LVC. Specifically, we classify merging systems as follows: `BNS' for both $M_1$ and $M_2$ less than $M_{{\rm{NS}}}^{\max }$, `BH--NS' for $M_1>5M_{\odot}$ and $M_2<M_{{\rm{NS}}}^{\max }$, `Gap--NS' for $M_{{\rm{NS}}}^{\max }<M_1<5M_{\odot}$ and $M_2<M_{{\rm{NS}}}^{\max }$. In this paper, the minimum NS mass is set to $1M_{\odot}$. 

Adopting the method discussed above, we derive the ejecta masses for BNS, BH-NS and Gap-NS mergers with given chirp masses. The results are plotted in Figure \ref{fig:mass_result}, which shows the dynamical ejecta (left) and disk (right) masses for effective BH spins of 0, 0.25, 0.5 and 0.75, respectively. The dashed lines for given chirp masses are shown for the cases that there is no NS disruption by BH (i.e., no dynamical ejecta or disk). 

As shown in Figure \ref{fig:mass_result}, the masses of disk and dynamical ejecta can be up to 0.3$M_{\odot}$ and 0.1$M_{\odot}$ for the SFHo EOS, respectively. The mass of disk is always larger than that of dynamical ejecta for all cases. For a GAP-NS or BH-NS merger of a given chirp mass, the disruption of NS is more likely to occur in the system with higher effective spin, lower NS mass or larger BH mass. Meanwhile, the amount of either type of ejecta increases with the initial effective spin of the merging system. 

As the fitting formulae that we used for BNS mergers do not depend on the spin of the system (see section 3.1.1 and 3.1.2), the masses of either dynamical ejecta or disk for those mergers presented in Figure 1 are independent of the spin. It is worth noting that for a given chirp mass in the range of $1.5M_{\odot}-1.7M_{\odot}$, the disk mass of a BNS merger is far lower than that of a GAP-NS merger. Besides, all the BNS mergers in those ranges of chirp masses hardly eject dynamical ejecta while some of the GAP-NS mergers eject dynamical ejecta (around 0.02$M_{\odot}$). 

In the next three sections, we use the results of ejecta given above to estimate the emissions of kilonova, sGRB, and cocoon. Throughout the paper we assume $50\%$ of the disk is driven out as disk wind for all the merger events.

%%%%%%%%%%%%%%%%%%%%%%%%%%%%%%%%%%%%%%%%%%%%%figure-2
\begin{figure}[tb]
\begin{center}
\includegraphics[width=8.7cm, angle=0]{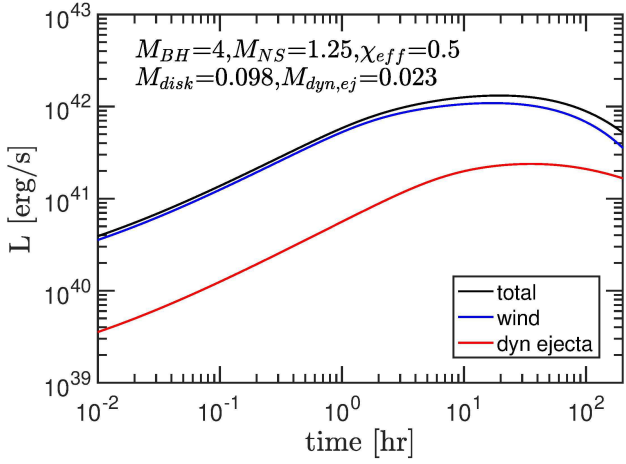}

\caption{Bolometric luminosity light curve for a kilonova from a merger with $M_{BH}=4M_{\odot}$, $M_{NS}=1.25M_{\odot}$ and $\chi_{eff}=0.5$. The velocity of the dynamical ejecta and disk wind to be 0.3$c$ and 0.2$c$ respectively. The opacity of the dynamical ejecta and disk wind is set to 10 $\rm cm^2 g^{-1}$ and 1 $\rm cm^2 g^{-1}$ respectively.}\label{fig:LC_KN}
\end{center}
\end{figure}
%%%%%%%%%%%%%%%%%%%%%%%%%%%%%%%%%%%%%%%%%%%%%figure-2

%%%%%%%%%%%%%%%%%%%%%%%%%%%%%%%%%%%%%%%%%%%%%%%%%%%%%%%%%%%%%%%%%%%%%%%%%%%%
\section{Kilonova}
In order to investigate the dependence of kilonova emission on the parameters of the merging system, we adopt a toy model of kilonova presented in \cite{metzger17} and apply it to the ejecta mass distribution with given chirp masses as shown in Figure  \ref{fig:mass_result}. 

\cite{rosswog14} performed simulations that focused on the the long-term evolution of the dynamical ejecta of a BNS merger and found that all remnants expand in a nearly homologous manner. Therefore, we assume that the merger ejecta is homologously expanding. The mean radius of a layer of ejecta with velocity $v$ is $R \approx vt$ at time $t$ following the merger. According to \cite{bauswein13}, a power-law function can be used to describe the distribution of ejecta mass with velocity greater than $v_0$ \citep{metzger17}
\begin{equation}
{M_v} = M(>v) = \begin{array}{*{20}{c}}
{M{{\left( {{v \mathord{\left/
 {\vphantom {v {{v_0}}}} \right.
 \kern-\nulldelimiterspace} {{v_0}}}} \right)}^{ - \beta }}}
\end{array}
\end{equation}
where $M$ is the total mass of ejecta, $v_0$ is the average ($\sim$ minimum) velocity, and the value of $\beta$ is set to 3. 

%%%%%%%%%%%%%%%%%%%%%%%%%%%%%%%%%%%%%%%%%%%%%%%%%%%%%%figure_3

\begin{figure*}[tb]
  \centering
    \subfigure{
    \begin{minipage}[t]{8.6cm}
    \includegraphics[width=8.6cm]{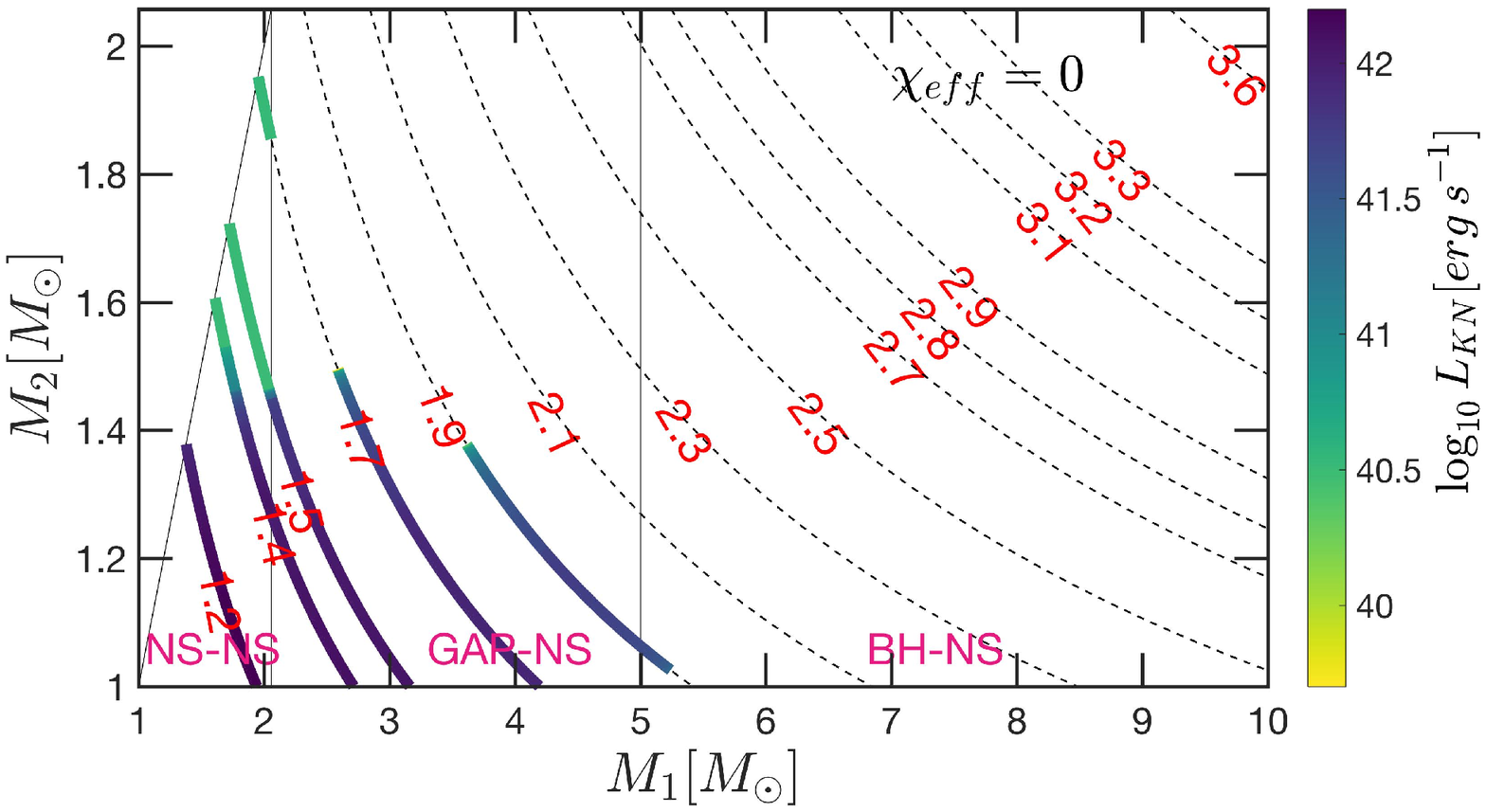}
    \end{minipage}} 
    \subfigure{
    \begin{minipage}[t]{8.6cm}
    \includegraphics[width=8.6cm]{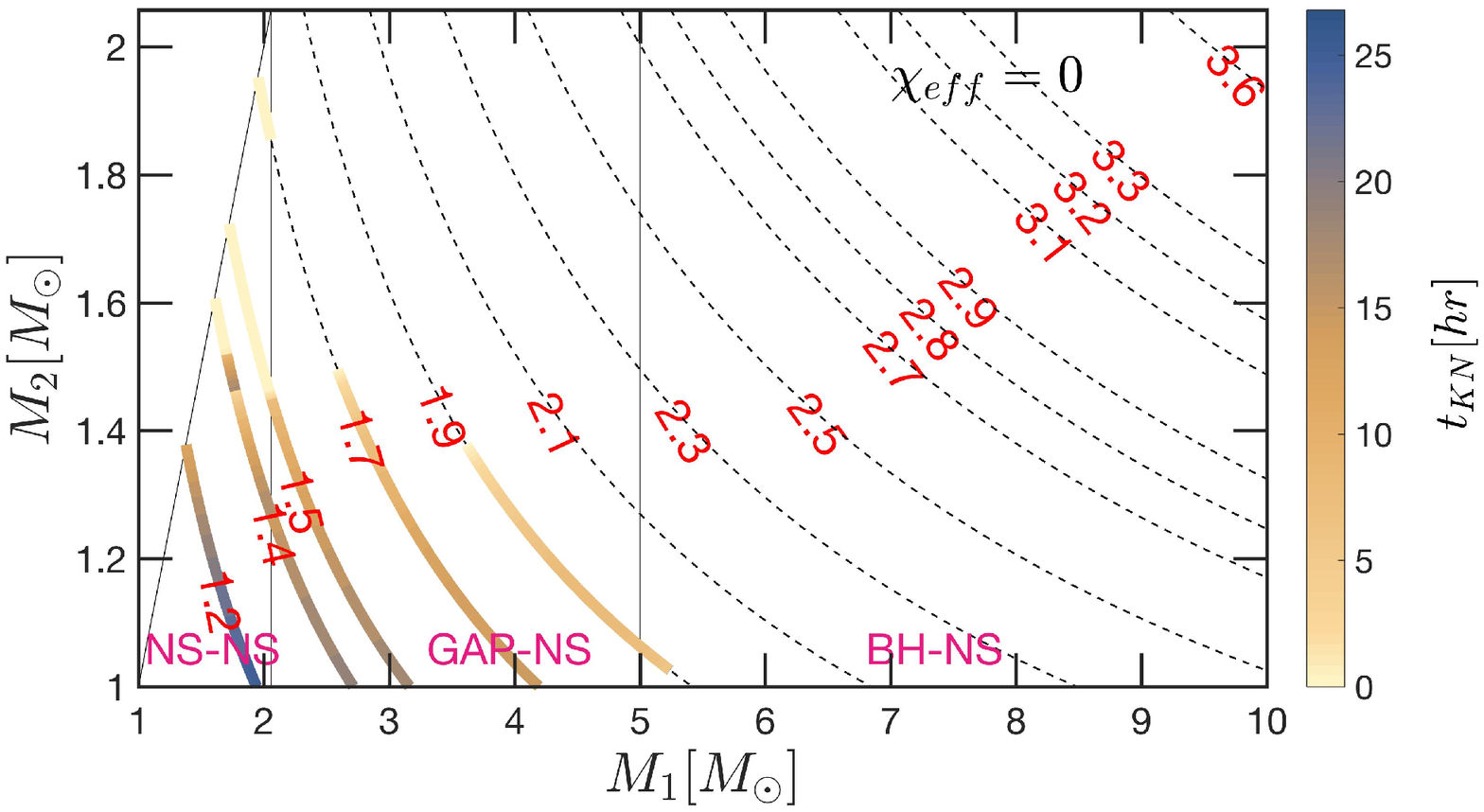}
    \end{minipage}} 
    
    \subfigure{
    \begin{minipage}[t]{8.6cm}
    \includegraphics[width=8.6cm]{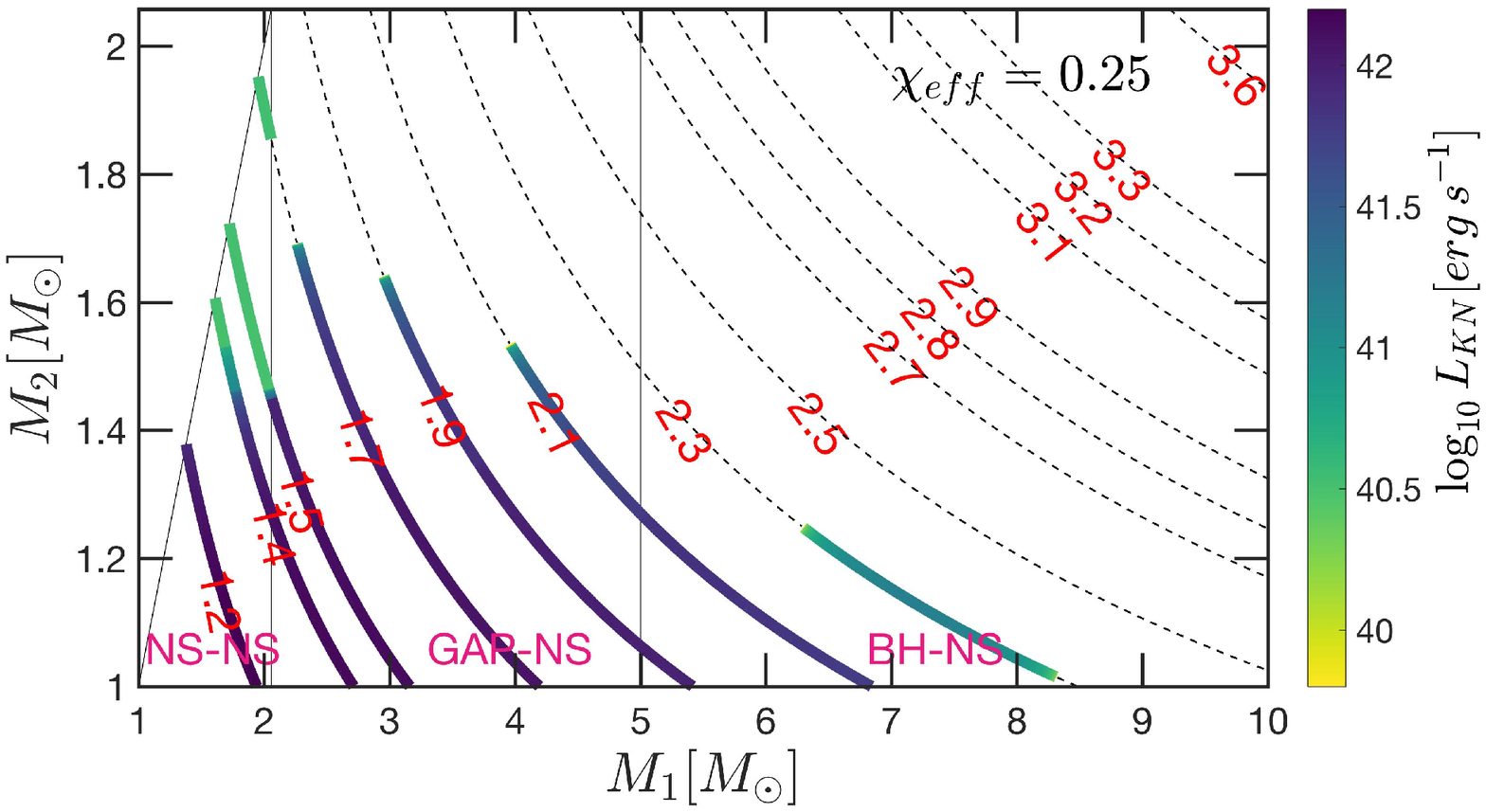}
    \end{minipage}} 
    \subfigure{
    \begin{minipage}[t]{8.6cm}
    \includegraphics[width=8.6cm]{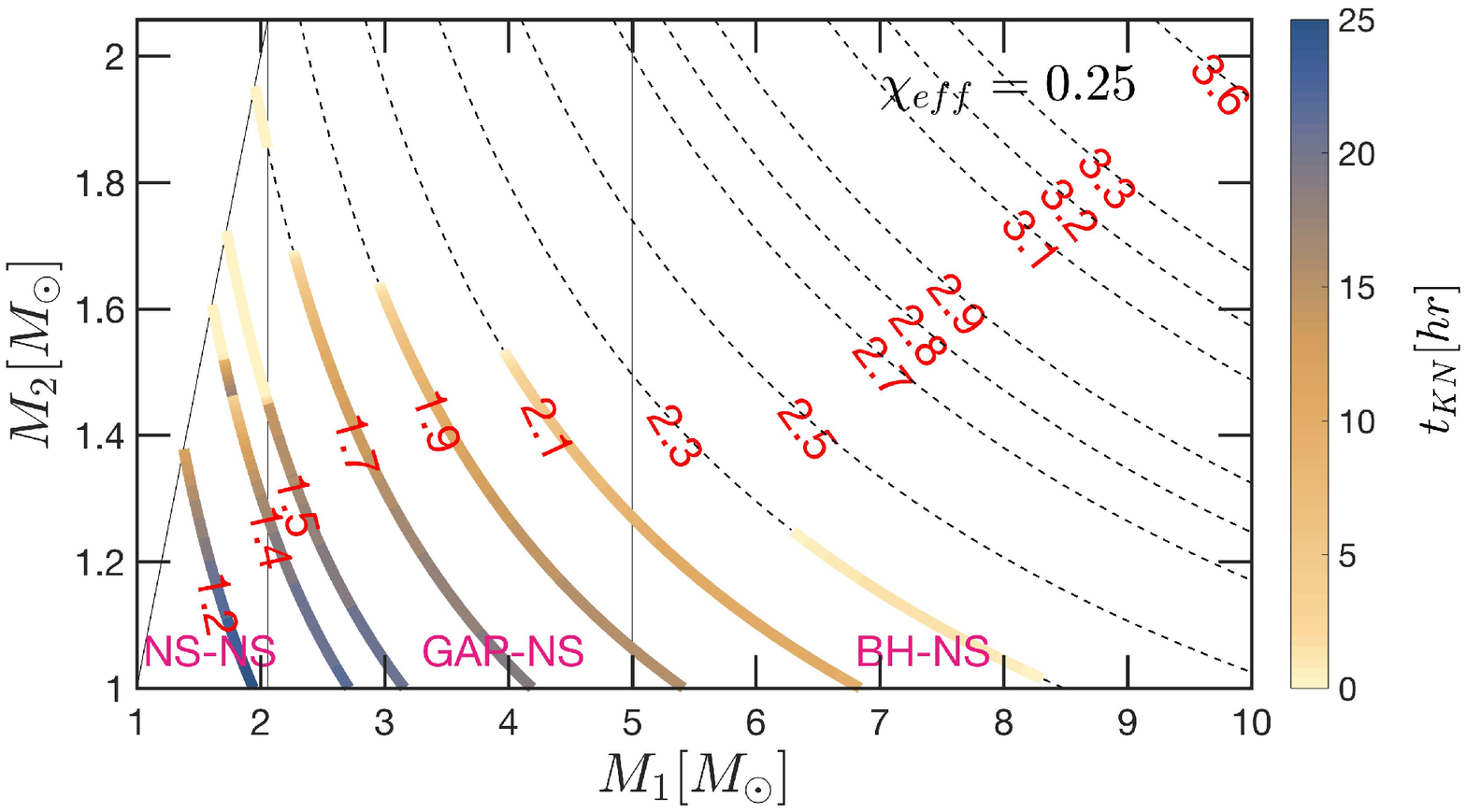}
    \end{minipage}} 
    	
    \subfigure{
    \begin{minipage}[t]{8.6cm}
    \includegraphics[width=8.6cm]{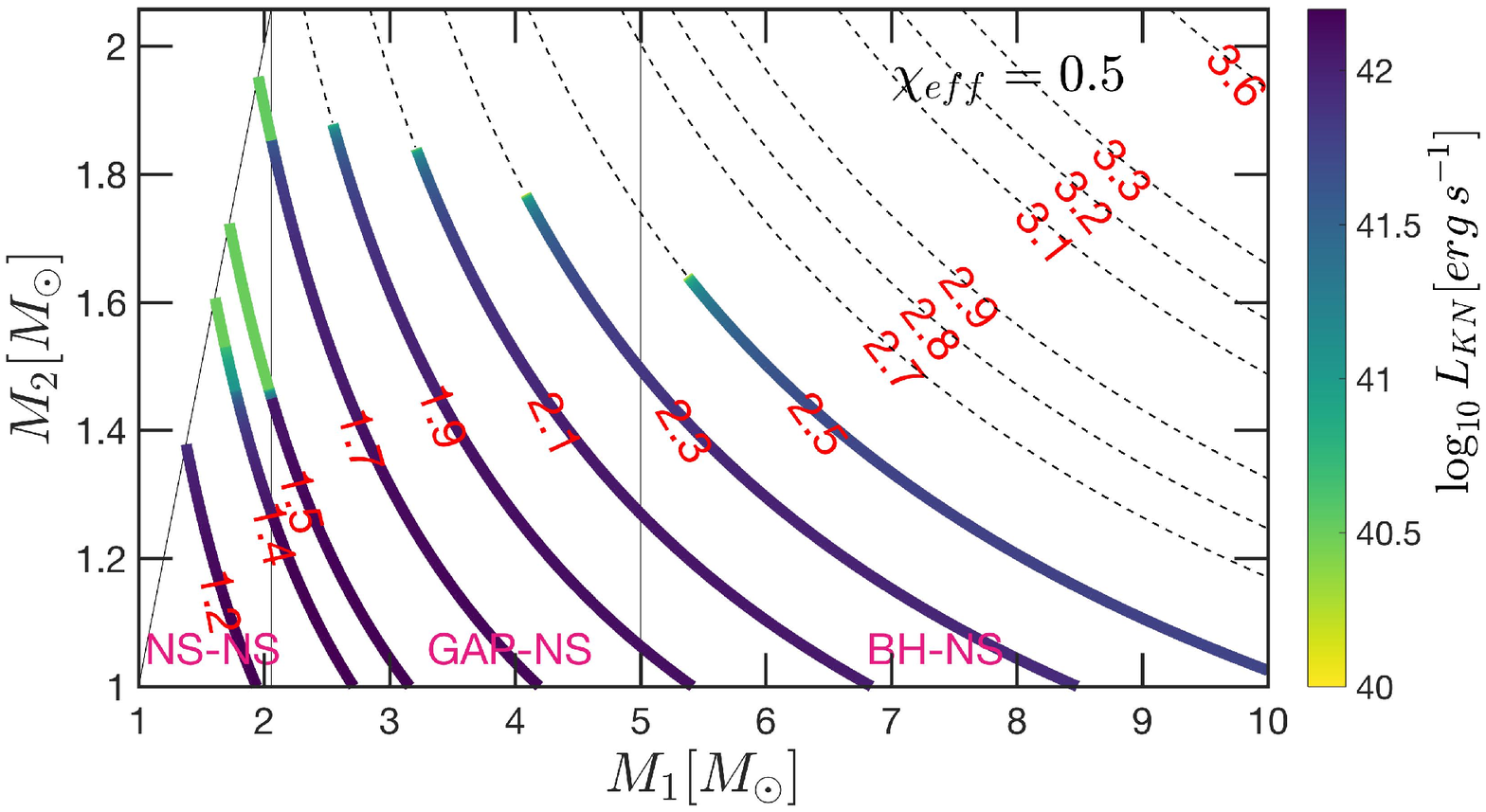}
    \end{minipage}} 
    \subfigure{
    \begin{minipage}[t]{8.6cm}
    \includegraphics[width=8.6cm]{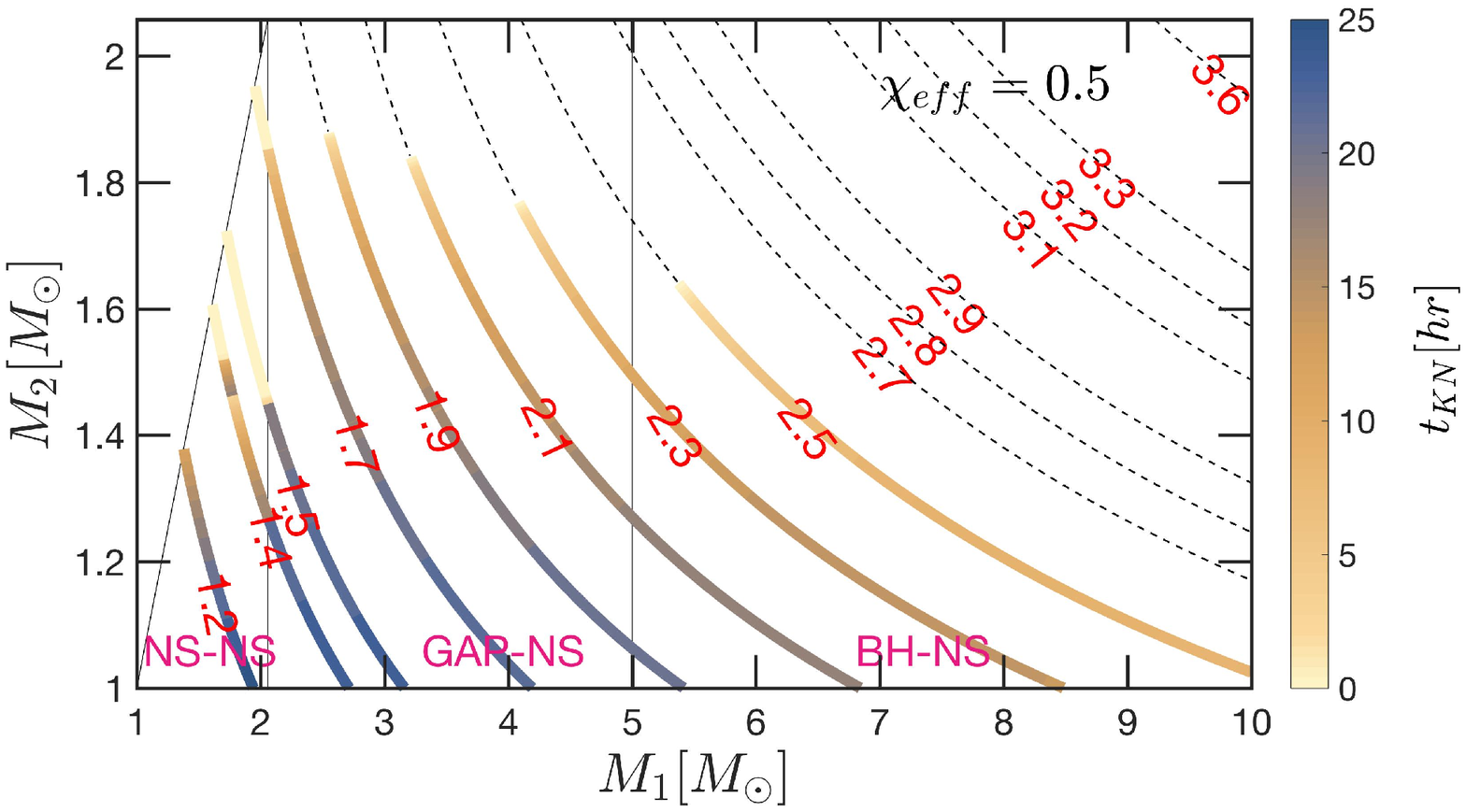}
    \end{minipage}} 
    
    \subfigure{
    \begin{minipage}[t]{8.6cm}
    \includegraphics[width=8.6cm]{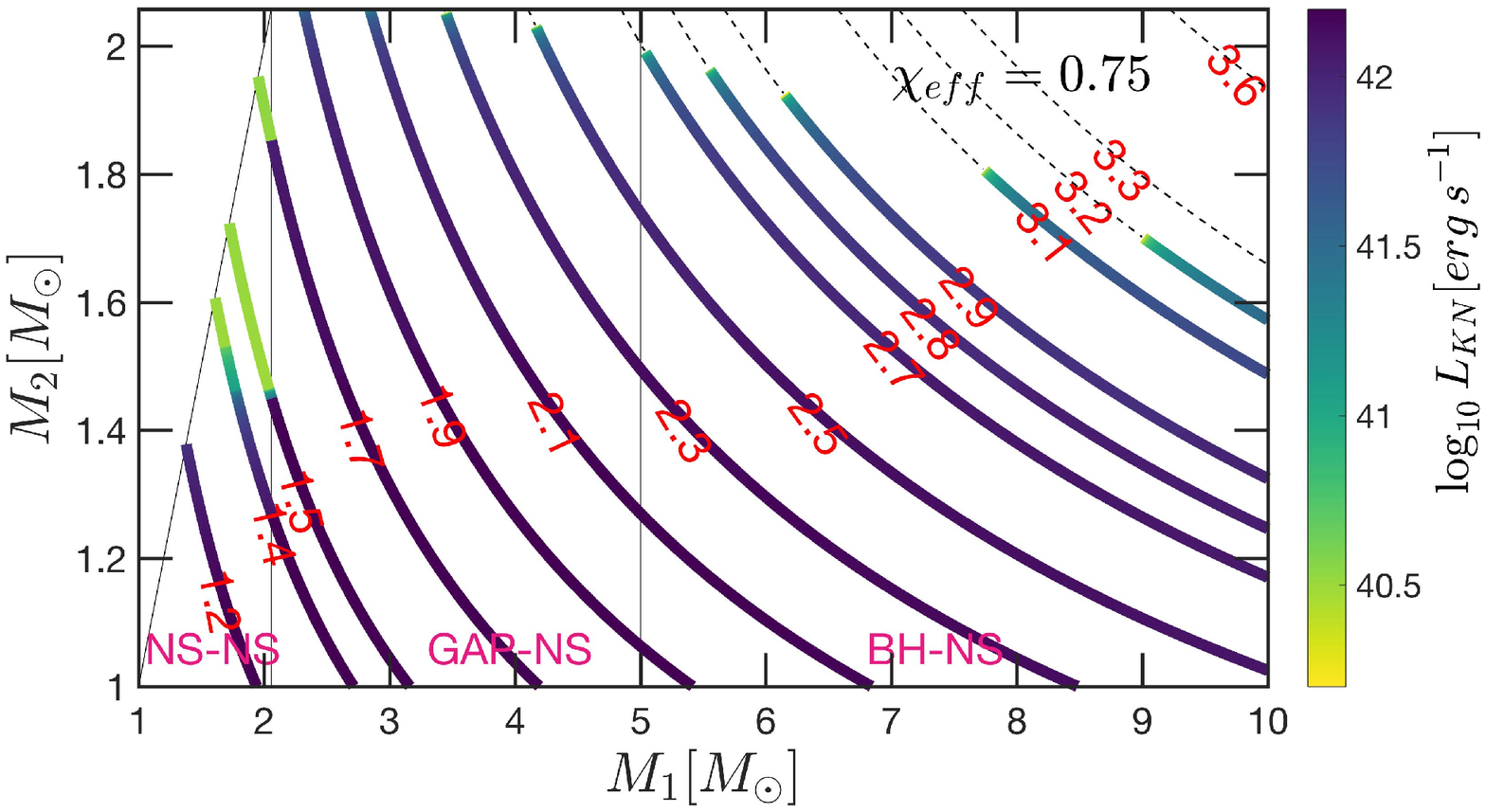}
    \end{minipage}} 
    \subfigure{
     \begin{minipage}[t]{8.6cm}
    \includegraphics[width=8.6cm]{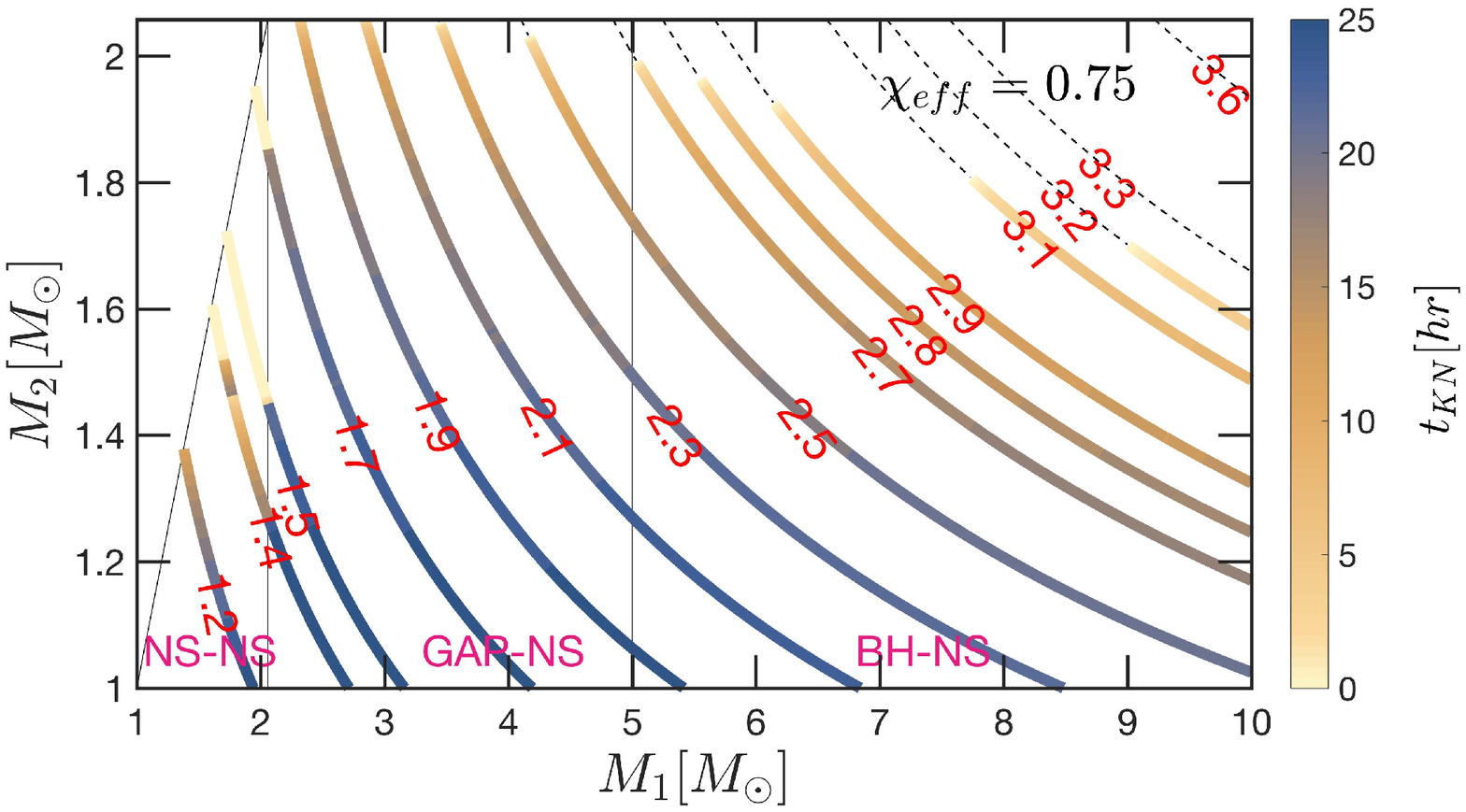}
    \end{minipage}}    
    \centering 
   \caption{Peak luminosity (left) and peak time (right) of kilonova for given chirp mass $M_{ch}$ whose values are shown in red numbers. From top to bottom, rows of panels correspond to effective BH spin of 0, 0.25, 0.5 and 0.75, respectively. The velocities of the dynamical ejecta and disk wind are set to be 0.3c and 0.2c, respectively. The opacities of the dynamical ejecta and disk wind are set to 10 $\rm cm^2 g^{-1}$ and 1 $\rm cm^2 g^{-1}$, respectively.}\label{fig:KN_result}
\end{figure*}

%%%%%%%%%%%%%%%%%%%%%%%%%%%%%%%%%%%%%%%%%%%%%%%%%%%%%figure_3 

Initially, the ejecta is extremely hot. However, this thermal energy cannot escape as radiation due to the high optical depth of ejecta at early time. The optical depth is given by the following equation as
\begin{equation}
\begin{split}
&\tau  = \rho \kappa R = \frac{{3M\kappa }}{{4\pi {R^2}}}\\ 
&\simeq 70\left( {\frac{M}{{{{10}^{ - 2}}{M_ \odot }}}} \right)\left( {\frac{\kappa }{{1\;{\rm{c}}{{\rm{m}}^{\rm{2}}}{{\rm{g}}^{{\rm{ - 1}}}}}}} \right){\left( {\frac{v}{{0.1c}}} \right)^{ - 2}}{\left( {\frac{t}{{1\;{\rm{day}}}}} \right)^{ - 2}}
\end{split}
\end{equation}
where $\rho  = {{3M} \mathord{\left/{\vphantom {{3M} {\left( {4\pi {R^3}} \right)}}} \right. \kern-\nulldelimiterspace} {\left( {4\pi {R^3}} \right)}}$ is the mean density and $\kappa$ is the opacity (cross section per unit mass). 

As the ejecta expands, the diffusion timescale decreases with time $t_{diff}$ $\propto$ $t^{-1}$, and eventually radiation escapes once $t_{diff}$ = $t$ \citep{arnett82}. The diffusion timescale of the whole ejecta is
\begin{equation}
{t_{{{diff}}}} \simeq \frac{R}{c}\tau  = \frac{{3M\kappa }}{{4\pi cR}} = \frac{{3M\kappa }}{{4\pi cvt}}.
\end{equation}
For the individual mass layer $M_v$ the radiation escapes on the diffusion timescale
\begin{equation}
{t_{d,v}} \approx \frac{{3{M_v}{\kappa _v}}}{{4\pi \beta {R_v}c}}\mathop  = \limits_{{R_v} = vt} \frac{{M_v^{{4 \mathord{\left/
 {\vphantom {4 3}} \right.
 \kern-\nulldelimiterspace} 3}}{\kappa _v}}}{{4\pi {M^{{1 \mathord{\left/
 {\vphantom {1 3}} \right.
 \kern-\nulldelimiterspace} 3}}}{v_0}tc}}
\end{equation}
where $\kappa_v$ is the opacity of the mass layer $v$, and the second equality makes use of Eq. (6) with $\beta$ = 3. 

For the layer whose differential mass is $\delta M_v$, its location evolves as 
\begin{equation}
\frac{{d{R_v}}}{{dt}} = v.
\end{equation}
Its thermal energy $\delta E_v$ evolves as
\begin{equation}
\frac{{d\left( {\delta {E_v}} \right)}}{{dt}} =  - \frac{{\delta {E_v}}}{{{R_v}}}\frac{{d{R_v}}}{{dt}} - {L_v} + \dot Q
\end{equation}
where the first term accounts for losses due to PdV expansion in the radiation-dominated ejecta. The second term
\begin{equation}
{L_v} = \frac{{\delta {E_v}}}{{{t_{d,v}} + {t_{lc,v}}}}
\end{equation}
is the observed luminosity for each mass layer and $t_{lc, v}$ = $R_v/c$ serves as a lower limit to the energy loss time. 

The third term $\dot{Q}_{r, v}$ accounts for radioactivity heating by the radioactive decay of heavy r-process nuclei and can be expressed as
\begin{equation}
{{\dot Q}_{r,v}} = \delta {M_v}{X_{r,v}}{{\dot e}_r}\left( t \right)
\end{equation}
where $X_{r,v}$ is the r-process mass fraction in mass layer $M_v$ and $\dot e_r$ is the specific heating rate. Although there is a small amount (up to $\sim$ 10\%) of ejecta expands sufficiently rapidly to allow most neutrons to avoid being captured into nuclei \citep{metzger15,just15,mendoza15,fernandez19}, most of the ejecta follows a relatively slow expansion and is dense enough to cause the capture of all neutrons via the r-process \citep{metzger15,just15,mendoza15}. In this work, we adopt $X_{r,v}=1$ for simplicity. For neutron-rich ejecta, $\dot e_r$ can be approximated by the following formula \citep{korobkin12}
\begin{equation}
\begin{split}
{{\dot e}_r} = &4 \times {10^{18}}{\rm{erg}}\,{{\rm{s}}^{ - 1}}\,{{\rm{g}}^{ - 1}}\\
&\times {\epsilon_{{\rm{th,}}v}}{\left( {0.5 - {\pi ^{ - 1}}\arctan [{{\left( {t - {t_0}} \right)} \mathord{\left/
 {\vphantom {{\left( {t - {t_0}} \right)} \sigma }} \right.
 \kern-\nulldelimiterspace} \sigma }]} \right)^{1.3}}
\end{split}
\end{equation}
where $t_0$ = 1.3 s, $\sigma$ = 0.11 s, and $\varepsilon_{th,v}$ is the thermalization efficiency (see below).

The value of the thermal efficiency $\varepsilon_{th,v}$ decreases from 0.5 to 0.1 during the first week since the merger \citep{barnes16}.  In what follows, we adopt the fit provided in \cite{barnes16},
\begin{equation}
{\epsilon_{{\rm{th,}}v}}\left( t \right) = 0.36[exp\left( { - {a_v}{t_{{\rm{day}}}}} \right) + \frac{{\ln \left( {1 + 2{b_v}t_{{\rm{day}}}^{{c_v}}} \right)}}{{2{b_v}t_{{\rm{day}}}^{{c_v}}}}]
\end{equation}
where $t_{day}$ = $t$/(1 day). According to \cite{metzger17}, we adopt fixed values of $a_v$ = 0.56, $b_v$ = 0.17 and $c_v$ = 0.74.

Numerically solving Eqs. (11) and (12) for each layer  gives $L_v$. Summing $L_v$ over all mass shells gives the total luminosity
\begin{equation}
{L_{tot}} \simeq \sum\nolimits_v {L_v}. 
\end{equation}

In our calculation we set the velocity of the dynamical ejecta and disk wind to be 0.3$c$ and 0.2$c$, respectively. The opacities of the dynamical ejecta and disk wind are approximated to be 10 $\rm cm^2 g^{-1}$ and 1 $\rm cm^2 g^{-1}$, respectively \citep{kasen17}.

Figure \ref{fig:LC_KN} shows the bolometric luminosity light curve for a kilonova from a merger with $M_{BH}=4M_{\odot}$, $M_{NS}=1.25M_{\odot}$ and $\chi_{eff}=0.5$, which produces ejecta with $M_{wind}=0.049M_{\odot}$ and $M_{dyn}=0.023M_{\odot}$. Figure \ref{fig:KN_result} illustrates the variation of peak luminosity (left) and peak time (right) of kilonova versus the parameters of merging system. Note that the peak timescales illustrated in Figure \ref{fig:KN_result} corresponds to the time at which the total luminosity contributed from both wind and dynamical ejecta reaches its maximum value.  

As shown in Figure \ref{fig:KN_result}, the peak luminosities range from $10^{40}$ to $10^{42.5}$ erg/s, and the peak timescales are $\leq$ 25 hrs. The comparison between Figure \ref{fig:mass_result} and Figure \ref{fig:KN_result} shows that the peak timescales increase with the total  masses of disk wind and dynamical ejecta. Besides, Figure \ref{fig:KN_result} indicates a positive dependence of the peak timescales on $\chi_{eff}$, as a merger system with higher spin would eject larger amount of dynamical ejecta and disk wind. For most of the GAP-NS mergers, the peak timescale of the kilonova emission is relatively large (with a median value of $\sim$ 20 hrs, see the right column of Figure \ref{fig:KN_result}) due to the sufficient ejection of both dynamical ejecta and disk wind, thus it should be easier to detect its emission. 

As we have shown in section 3.2, for a given chirp mass at the range of $1.5M_{\odot}-1.7M_{\odot}$, all the BNS mergers eject negligible dynamical ejecta and relatively massive disk wind, while the masses of both dynamical ejecta and disk wind ejected from GAP-NS mergers are larger than those of BNS mergers (see Figure \ref{fig:mass_result}). In addition, the opacity of disk wind is lower than that of dynamical ejecta \citep{kasen17}. Therefore, the kilonova emission of the BNS mergers with a given chirp mass at those ranges would be bluer and peak earlier than that of GAP-NS mergers (see Figure \ref{fig:KN_result}). This difference between the BNS merger and GAP-NS merger for a given chirp mass at the range of $1.5M_{\odot}-1.7M_{\odot}$ would be more significant if the initial effective BH spin is higher, and hence could be applied to identify a `mass gap' BH.

%%%%%%%%%%%%%%%%%%%%%%%%%%%%%%%%%%%%%%%%%%%%%%%%

%%%%%%%%%%%%%%%%%%%%%%%%%%%%%%%%%%%%%%%%%%%%%%%%%%%%%%%%%%%%%%%%%%%%%%%%%%%%
\section{sGRB}
In this section we estimate the luminosity and time scale of the sGRB, and their dependence on the binary parameters. If the merger remnant is a rotating BH with a massive disk, a sGRB is probably powered by accretion onto the stellar mass black hole. Due to the large accretion rate, the BH can spin up rather quickly. If a strong magnetic field threads the spinning BH and is connected with an external astrophysical load, the BH spin energy can be tapped via the Blandford--Znajek mechanism \citep{blandford77} which launches a Poynting-flux-dominated jet.

The duration of emission in gamma/X-ray bands from the sGRB jet is determined by the characteristic viscous time-scale for accretion of the remnant disk/torus on to the BH \citep{fernandez15a, fernandez16}:
\begin{equation}
\begin{split}
{t_\gamma } &= {t_{{\rm{visc}}}} = \frac{{{r^2}}}{\nu_\alpha} \\
&\approx 0.26{\left( {\frac{\alpha }{{0.1}}} \right)^{ - 1}}{\left( {\frac{r}{{30\;{\rm{km}}}}} \right)^{{3 \mathord{\left/
 {\vphantom {3 2}} \right.
 \kern-\nulldelimiterspace} 2}}}{\left( {\frac{{{M_{{\rm{BH}}}}}}{{3{M_ \odot }}}} \right)^{{{ - 1} \mathord{\left/
 {\vphantom {{ - 1} 2}} \right.
 \kern-\nulldelimiterspace} 2}}}{\left( {\frac{H}{{0.1r}}} \right)^{ - 2}}\;{\rm{s}}
\end{split}
\end{equation}
where $r$ is the radial extent of the torus in which most of its mass and angular momentum is concentrated, $\nu_{\alpha}=\alpha{c_s}H$ is the disk viscosity due to turbulence \citep{shakura73}, $\alpha$ is the viscosity parameter, $c_s$ is the sound speed within the disk, and $H \approx c_s(GM_{BH}/r^3)^{-1/2}$ is the vertical scale height of the disk. The ratio of the disk scale height to the disk radius $H/r$ is assumed to be 0.1 for all the mergers in this work. Note that we set the mass of the promptly formed BH to be $M_1 + M_2 - M_{disk} - M_{dyn}$, ignoring the loss by GWs, for simplicity. 

The total power of Poynting flux from the BZ process can be estimated as \citep{lee00, li00, wang02, lei05,lei13, mckinney05, lei11}
\begin{equation}
\begin{split}
{L_{{\rm{BZ}}}} = &1.7 \times {10^{50}}\\
&\times{\chi _{{\rm{BH}}}^2}{\left( {\frac{{{M_{{\rm{BH}}}}}}{{1{M_ \odot }}}} \right)^2}\left( {\frac{B}{{{{10}^{15}}G}}} \right)^2f\left( {{\chi _{{\rm{BH}}}}} \right)\;{{{\rm{erg}}} \mathord{\left/
 {\vphantom {{{\rm{erg}}} {\rm{s}}}} \right.
 \kern-\nulldelimiterspace} {\rm{s}}}
\end{split}
\end{equation}
where $\chi_{BH}$ is the dimensionless spin parameter of the BH, and 
\begin{equation}
f\left( {{\chi _{{\rm{BH}}}}} \right) = \left[ {{{\left( {1 + {q^2}} \right)} \mathord{\left/
 {\vphantom {{\left( {1 + {q^2}} \right)} q}} \right.
 \kern-\nulldelimiterspace} q}} \right]\left[ {\left( {q + {1 \mathord{\left/
 {\vphantom {1 q}} \right.
 \kern-\nulldelimiterspace} q}} \right)\arctan q - 1} \right], 
\end{equation}
here $q = {{{\chi _{{\rm{BH}}}}} \mathord{\left/
 {\vphantom {{{\chi _{{\rm{BH}}}}} {\left( {1 + \sqrt {1 - \chi _{{\rm{BH}}}^2} } \right)}}} \right.
 \kern-\nulldelimiterspace} {\left( {1 + \sqrt {1 - \chi _{{\rm{BH}}}^2} } \right)}}$. 
 
%%%%%%%%%%%%%%%%%%%%%%%%%%%figure-4
\begin{figure}[t]
\begin{center}
\includegraphics[width=8cm, angle=0]{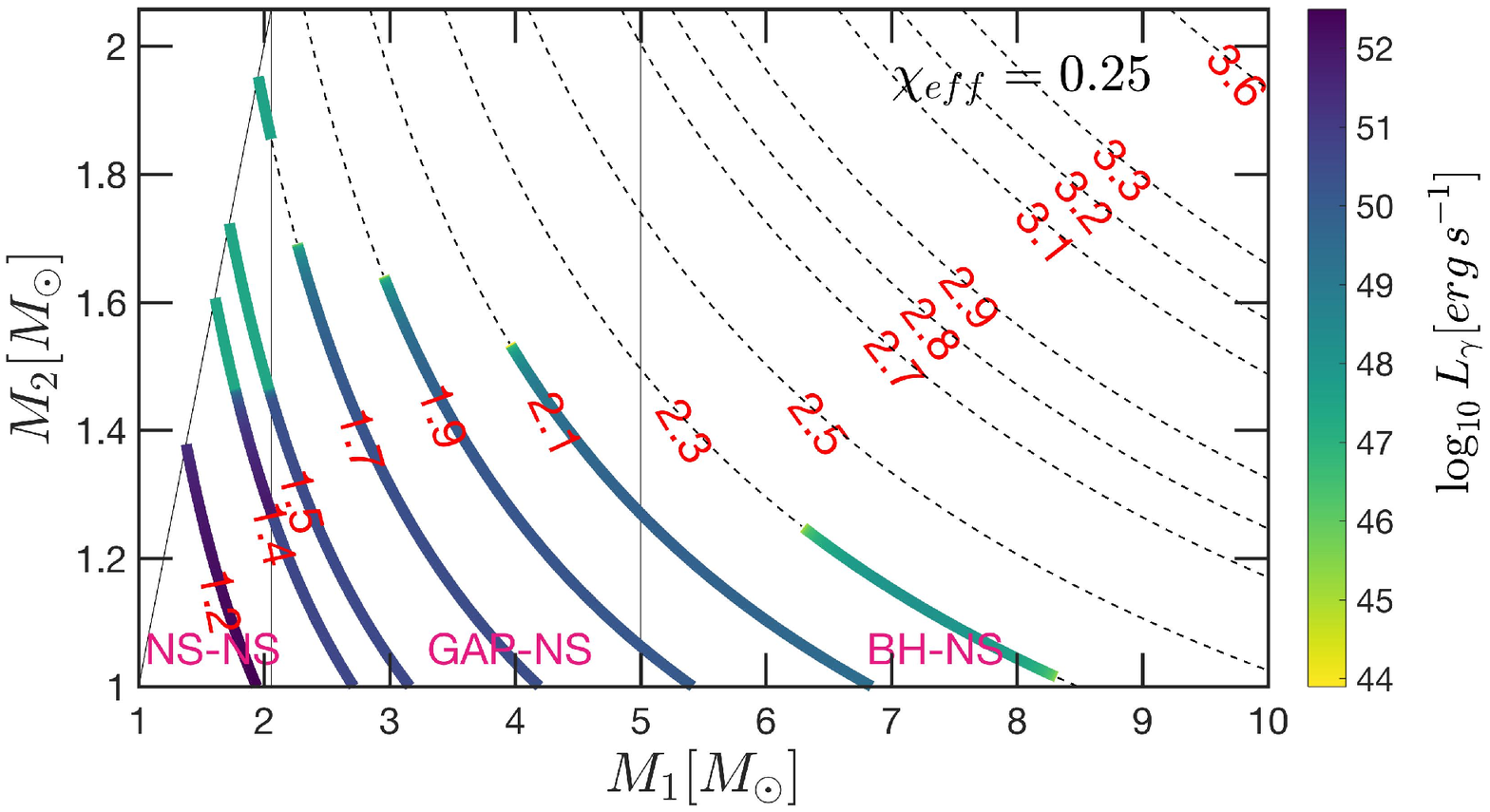}
\includegraphics[width=8cm, angle=0]{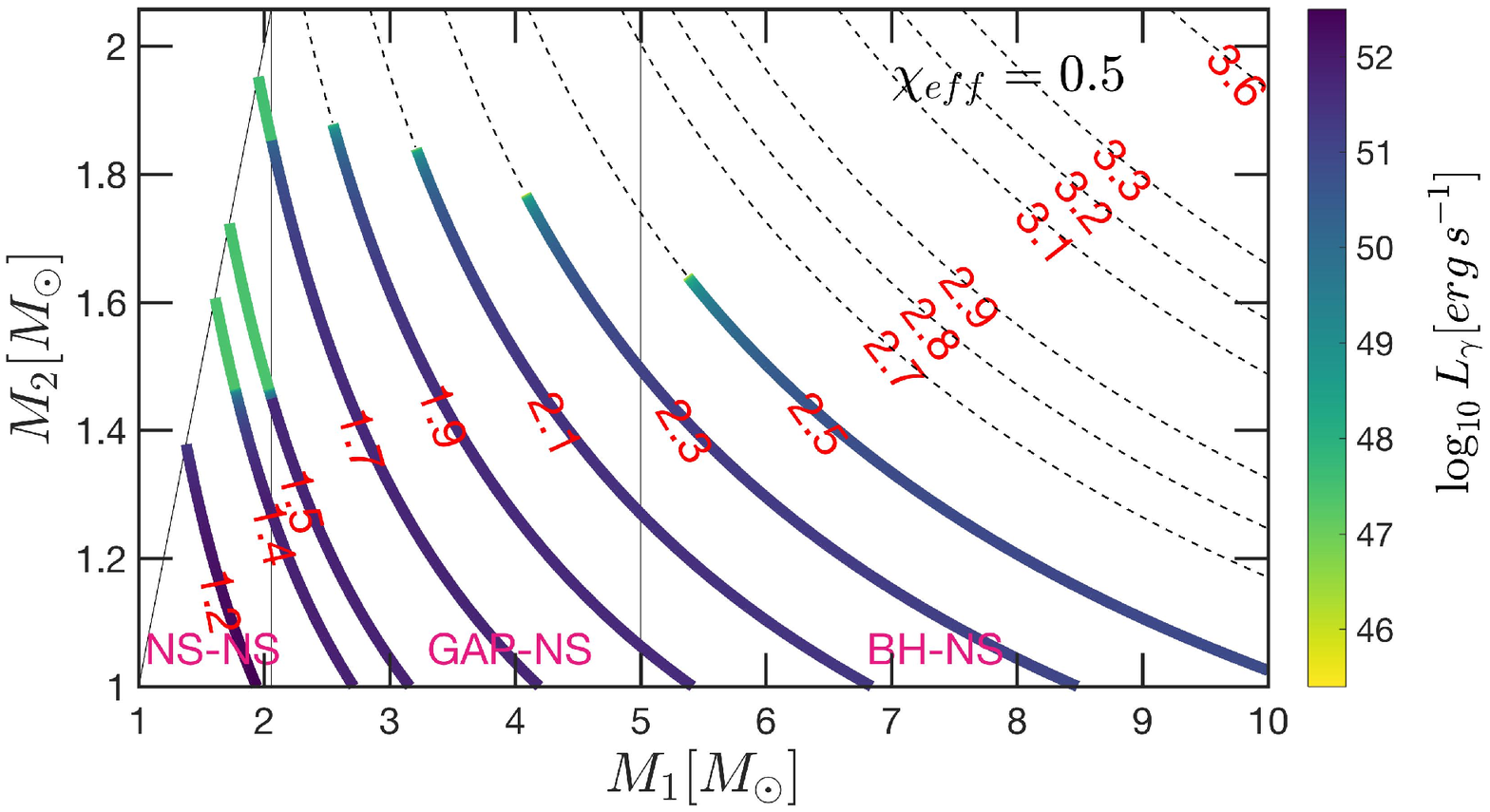}
\includegraphics[width=8cm, angle=0]{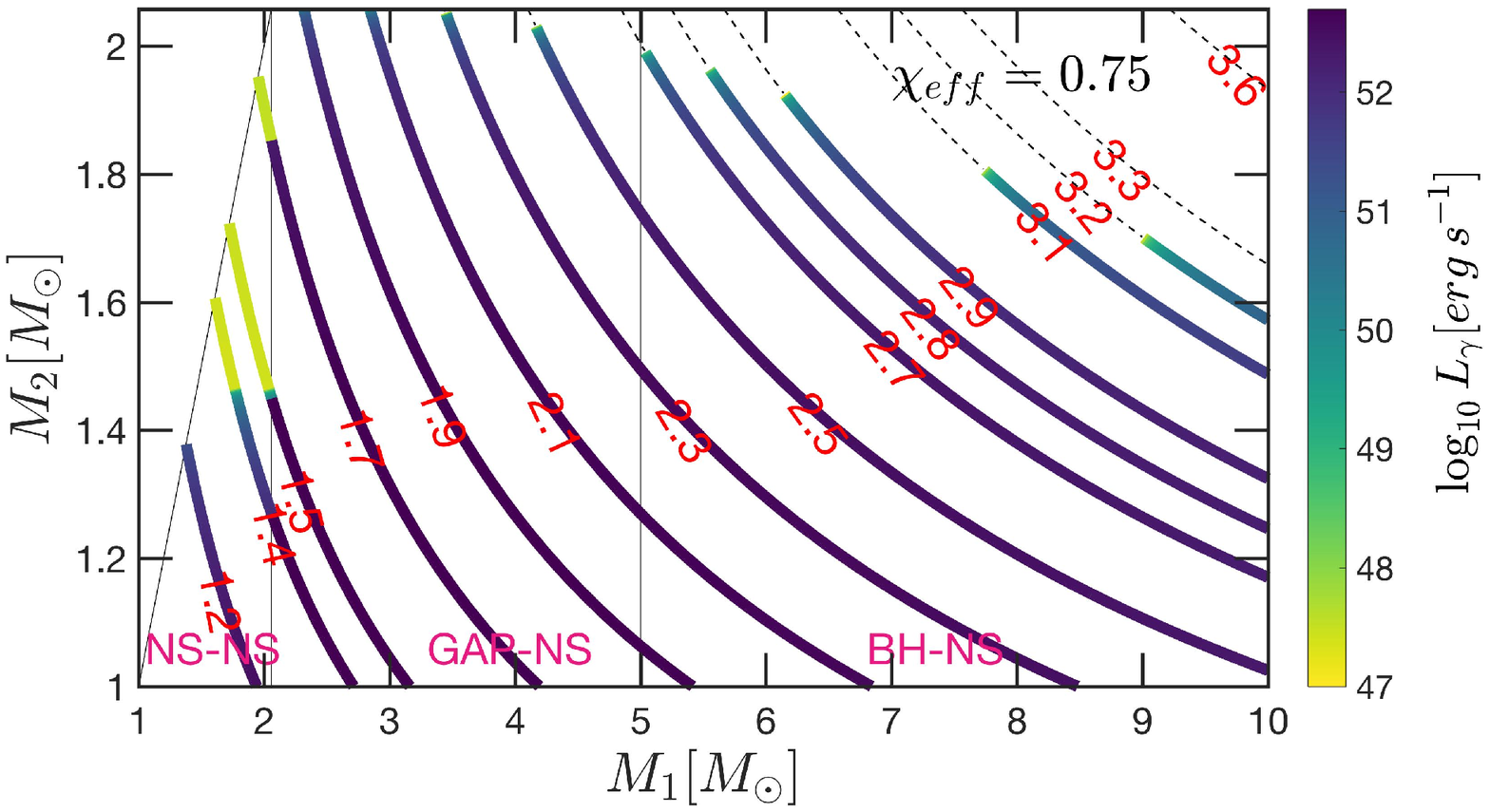}
\caption{The peak isotropic equivalent gamma-rays luminosity for sGRB prompt emission corresponding to initial BH spin of 0.25 (top), 0.5 (middle) and 0.75 (bottom). Note that for the BNS mergers whose $M_{ch}=1.2,1.4,1.5$ and 1.7$M_{\odot}$, the masses of the remnant objects are all larger than $M_{{\rm{NS}}}^{\max }$, thus we consider those BNS merger remnants as BHs. The result for BNS mergers corresponds to the situation that a BH formed promptly.}\label{fig:LpeakGRB}
\end{center}
\end{figure}
%%%%%%%%%%%%%%%%%%%%%%%%%%%figure-4

As shown in Eq. (18), $L_{BZ}$ depends on $M_{BH}$, $\chi_{BH}$ and $B$. A strong magnetic field of the order $10^{15} G$ is required to produce the high luminosity of a sGRB. The accumulation of magnetic flux from an accretion flow may account for such a high magnetic field strength \citep{tchekhovskoy11}. 

Considering the balance between the magnetic pressure on the horizon and the ram pressure of the innermost part of the accretion flow \citep{moderski97}, one can estimate the magnetic field strength threading the BH horizon as
\begin{equation}
B \simeq 7.4 \times {10^{16}}{{\dot M}^{{1 \mathord{\left/
 {\vphantom {1 2}} \right.
 \kern-\nulldelimiterspace} 2}}}{(\frac{M_{{\rm{BH}}}}{1M_\odot})}^{ - 1}{\left( {1 + \sqrt {1 - \chi _{{\rm{BH}}}^2} } \right)^{ - 1}} G.
\end{equation}
Assuming the accretion time is equal to $t_{visc}$, the accretion rate of BH is
\begin{equation}
\dot{M} = {\frac{(1 - \xi_{wind})\times M_{disk}}{t_{visc}}} 
\end{equation}
where $\xi_{wind}$ accounts for the disk mass lost in the form of winds.

Substituting Eqs. (20) $\&$ (21) into Eq. (18), one can estimate the luminosity of $\gamma$-rays for a one-sided jet as
\begin{equation}
{L_{{\rm{\gamma}}}} = 4.5 \times {10^{53}} {\frac {\epsilon_\gamma \chi _{{\rm{BH}}}^2 (1 - \xi_{wind}){g}\left( {{\chi _{{\rm{BH}}}}} \right) M_{disk} }{t_{visc}}}\;{\rm{erg/s}}
\end{equation}
where ${g}\left( {{\chi _{{\rm{BH}}}}} \right) = {{f\left( {{\chi _{{\rm{BH}}}}} \right)} \mathord{\left/
 {\vphantom {{F\left( {{\chi _{{\rm{BH}}}}} \right)} {{{\left( {1 + \sqrt {1 - \chi _{{\rm{BH}}}^2} } \right)}^2}}}} \right.
 \kern-\nulldelimiterspace} {{{\left( {1 + \sqrt {1 - \chi _{{\rm{BH}}}^2} } \right)}^2}}}$. The efficiency factor $\epsilon_\gamma$ accounts for the conversion of Poynting-flux-dominated energy into $\gamma$-rays energy. Note that in this work, we set $\epsilon_\gamma$ to be 0.015 \citep{barbieri19b}.

Considering a simple top-hat jet model, the isotropic equivalent $\gamma$-ray luminosity for the sGRB prompt emission is then given by
\begin{equation}
 L_{\gamma,iso}=L_\gamma/(1-cos\theta_j) \simeq L_{\gamma}/(\theta_j^2/2)
\end{equation}
where $\theta_{j}$ is the jet half-opening angle \citep{frail01}. It is worth noting that we assume the jet half-opening angle $\theta_j=0.1$ in this work.

Based on the above discussion for sGRB prompt emission, we calculated $L_{\gamma,iso}$ for three different BH spins of 0.25, 0.5 and 0.75. The results are shown in Figure \ref{fig:LpeakGRB}. Since the aforementioned calculation is constructed for the BZ jet from a rotating BH, it is necessary to identify whether a BNS merger forms a BH. In our calculation, for the BNS mergers whose $M_{ch}=1.2,1.4,1.5$ and $1.7M_{\odot}$, we consider the merger remnant to be a BH as long as the mass of the remnant object is larger than $M_{{\rm{NS}}}^{\max }$, and hence the results in Figure \ref{fig:LpeakGRB} for BNS mergers correspond to the situation that a BH forms promptly. The dimensionless spin parameter of the BH $\chi_{BH}$ is referred to the spin of the promptly formed BH. For the GAP-NS (BH-NS) merger, at the beginning of accretion, the spin of the BH is unchanged. Thus in our calculation, $\chi_{BH}$ is adopted to be the initial BH spin. For a non-spinning BNS, simulations show that the spin of the promptly formed BH ranges from 0.55 to 0.83 with a mean value of 0.68 \citep{coughlin19}, thus here we adopt the spin of the BH to be 0.68 in all cases for simplicity.

As illustrated in Figure \ref{fig:LpeakGRB}, for the sGRB powered by accretion onto the BH, the isotropic equivalent luminosity of $\gamma$-rays varies from $10^{45}$ erg/s to $10^{53}$ erg/s. The isotropic luminosities for most of the detected sGRBs are in the range of $10^{49}-10^{52}$ erg/s, while there are some abnormally low events like GRB170817A whose $L_{\gamma,iso}=(1.6\pm0.6)\times10^{47}$erg/s \citep{abbott17b,zhang18}. These observations are consistent with our predictions. Figure \ref{fig:LpeakGRB} suggests that low $L_{\gamma,iso}$ ($<10^{49}$erg/s) probably occurs in the BNS systems with the lighter one larger than $1.4M_{\odot}$, or in the BH--NS (GAP--NS) systems that involve a heavy NS. Besides, most of the BH--NS (GAP--NS) events probably have a $L_{\gamma,iso}$ higher than $10^{49}$erg/s. 

Due to the positive dependence of $L_{\gamma,iso}$ on the effective spin and the disk mass (see eqs. 22 \& 23; note that disk mass increases with the effective spin), a larger spin of the initial system (i.e., of the promptly formed BH) corresponds to a higher sGRB isotropic luminosity (see Figure \ref{fig:LpeakGRB}). For the GAP--NS (BH--NS) mergers with a fixed BH spin, the higher isotropic luminosities tend to occur in the systems with lower chirp masses. For a given chirp mass in the range of $M_{ch}=1.5M_{\odot}-1.7M_{\odot}$, as the disk masses of the BNS mergers are three orders of magnitude lower than that of GAP--NS mergers, the isotropic luminosities from BNS mergers are lower than those of GAP--NS mergers by around three to four orders of magnitude. This difference in $L_{\gamma,iso}$ for the chirp mass range of $1.5M_{\odot}-1.7M_{\odot}$ could be applied to identify a `mass gap' BH.

%%%%%%%%%%%%%%%%%%%%%%%%%%%%%%%%%%%%%%%%%%%%%%%%%%%%%%%%%%%%%%%%%%%%%%%%%%%%
\section{Cocoon}
When a relativistic jet propagates through the sub-relativistic material that was ejected during and after merger (hereafter referred as pre-burst ejecta), the interaction of the jet with the pre-burst ejecta would result in the formation of `cocoon', which is a highly-pressured, hot bubble enclosing the jet. If the jet breaks through those pre-brust ejecta successfully, the jet and cocoon together would form a structure which spreads over a much wider opening angle than that of the jet alone. The energy deposited by the shock in the cocoon diffuses as it expands and eventually escapes to the observer, producing an X-ray, UV and optical cooling emission \citep{nakar17, lazzati17}. The prompt emission from the cocoon is detectable for Swift BAT and Fermi GBM \citep{lazzati17}.  

\cite{lazzati17} presented a calculation of possible components of off-axis emission from a sGRB, including the cocoon prompt emission. Assuming a blackbody spectrum they obtain a cocoon emission luminosity
\begin{equation}
\begin{split}
{L_{\rm{c}}} &= 4 \times {10^{49}}erg/s\\
&\times{\left( {\frac{{{E_{\rm{c}}}}}{{{{10}^{49}erg}}}} \right)^{{2 \mathord{\left/
 {\vphantom {2 3}} \right.
 \kern-\nulldelimiterspace} 3}}}{\left( {\frac{{{\Gamma _{{\rm{c}}}}}}{{10}}} \right)^{{5 \mathord{\left/
 {\vphantom {5 3}} \right.
 \kern-\nulldelimiterspace} 3}}}{\left( {\frac{{{R_{\rm{a}}}}}{{{{10}^8cm}}}} \right)^{{{ - 1} \mathord{\left/
 {\vphantom {{ - 1} 3}} \right.
 \kern-\nulldelimiterspace} 3}}}
\end{split}
\end{equation}
where $E_c$, $\Gamma_c$ and $R_a$ correspond to the energy and Lorentz factor of the cocoon, and the distance that the jet head traverses (see Eq. 32), respectively. The prompt emission lasts for an angular time-scale as
 \begin{equation}
 \begin{split}
{t_{{\rm{c}}}} \sim {t_{{\rm{ang}}}} &= {{{R_{{\rm{ph,c}}}}} \mathord{\left/
 {\vphantom {{{R_{{\rm{ph,c}}}}} {\left( {c\Gamma _{\rm{c}}^2} \right)}}} \right.
 \kern-\nulldelimiterspace} {\left( {c\Gamma _{\rm{c}}^2} \right)}}\\
 &= 0.14\,{\rm{s}}{\left( {\frac{{{E_{\rm{c}}}}}{{{{10}^{49}}\,{\rm{erg}}}}} \right)^{{1 \mathord{\left/
 {\vphantom {1 2}} \right.
 \kern-\nulldelimiterspace} 2}}}{\left( {\frac{{\Gamma _{\rm{c}}}}{{{10}}}} \right)^{{{ - 7} \mathord{\left/
 {\vphantom {{ - 7} 2}} \right.
 \kern-\nulldelimiterspace} 2}}}
\end{split}
 \end{equation}
where ${R_{{\rm{ph}},{\rm{c}}}} = {\left( {\frac{{{E_c}{\sigma _T}}}{{8\pi {m_p}\Gamma _{{\rm{c}}}^3{c^2}}}} \right)^{{1 \mathord{\left/
 {\vphantom {1 2}} \right.
 \kern-\nulldelimiterspace} 2}}}$ is the photospheric radius of the cocoon \citep{lazzati17, bhattacharya19}, $\sigma_t$ is the Thompson cross-section, $m_p$ is the mass of proton. In the following two subsections, we consider the interaction of the jet with the pre-bust ejecta to estimate the jet breakout time $t_{b}$ and the Lorentz factor of cocoon $\Gamma_c$. Then we use them to determine the value of parameters in Eqs. (24) and (25), and eventually estimate $L_{c}$ and $t_{c}$ for the cocoon prompt emission. 
 
In the context of BH--NS mergers, the dynamical ejecta are mostly concentrated around the equatorial plane and the polar-shocked material is absent \citep{kyutoku13}. Nevertheless, the presence of disk winds and sufficient jet-launching time delay can result in the accumulation of the required ejecta mass around the polar regions. \cite{murguia20} found that the difference in the final structure of the jet is not highly sensitive to the exact structure of the wind. Thus in what follows, we consider only the disk wind as the pre-burst ejecta and assume the wind ejecta is isotropically distributed around the remnant, but bear in mind that this is a crude approximation to the real geometry of the pre-burst ejecta. 

%%%%%%%%%%%%%%%%%%%%%%%%%%%%%%%%%%%%%%%%%%%
%%%%%%%%%%%%%%%%%%%%%%%%%%%%%%%%%%%%%%%%%%%%
\subsection{Jet breakout time}

The interaction of the relativistic jet with the sub-relativistic pre-burst ejecta have been analytically studied by considering limiting cases for the dynamics of the ejecta, being either static \citep{begelman89, marti94, matzner03, bromberg11} or homologously expanding \citep{duffell18}. \cite{beniamini20} combine these two limits and describe the breakout time of a successful jet as
\begin{equation}
{t_{{\rm{b}}}} = {t_{{\rm{b,e}}}} + {t_{{\rm{b,s}}}}
\end{equation}
where $t_{b,e}$, $t_{b,s}$ are the jet breakout times in the homologous expansion and static ejecta limits, respectively.

In the static ejecta limit, the breakout time is the time that the jet takes to overpass the merger ejecta
\begin{equation}
{t_{{\rm{b,s}}}} = {t_{\rm{w}}}\frac{{{\beta _{{\rm{w}}}}}}{{{\beta _{\rm{h}}} - {\beta _{{\rm{w}}}}}}
\end{equation}
where $\beta_{w}$ is the velocity of the pre-brust ejecta (i.e., disk wind) and $\beta_h$ is the velocity of the jet's head, $t_w$ is the time interval between the merger and the launch of the jet. The velocity of the jet head is related to the ratio between the jet's total luminosity, $L_{j} \equiv L_{\gamma} / \epsilon_\gamma$, and the (isotropic equivalent) mass outflow rate of the wind ejecta, $\dot{M}_{w}$, as follows \citep{marti94, matzner03, bromberg11, murguia17} 
\begin{equation}
{\beta _{\rm{h}}} = \frac{{{\beta _{\rm{j}}} + {\beta _{{\rm{w}}}}{{\tilde L}^{{{ - 1} \mathord{\left/
 {\vphantom {{ - 1} 2}} \right.
 \kern-\nulldelimiterspace} 2}}}}}{{1 + {{\tilde L}^{{{ - 1} \mathord{\left/
 {\vphantom {{ - 1} 2}} \right.
 \kern-\nulldelimiterspace} 2}}}}}
\end{equation}
where
\begin{equation}
\begin{split}
\tilde L &\equiv \frac{{{L_{\rm{j}}}{\beta _{{\rm{w}}}}}}{{{{\dot M}_{{\rm{w}}}}{c^2}}} =\\
& \frac{{{0.14 }}}{{{\epsilon _\gamma }}}\left( {\frac{{{L_{{\rm{\gamma}}}}}}{{{{10}^{52}}{\rm{erg}}\,{{\rm{s}}^{ - 1}}}}} \right)\left( {\frac{{{\beta _{{\rm{w}}}}}}{{0.25}}} \right)\left( {\frac{{{{10}^{ - 2}}{M_ \odot }}}{{{M_{{\rm{w}}}}}}} \right)\left( {\frac{{{t_{\rm{w}}} + {t_{{\rm{b}}}}}}{{1\,{\rm{s}}}}} \right).
\end{split}
\end{equation}
As we have discussed above, in our calculation, $M_{w}$ is set to the total mass of disk wind for BH--NS merger. Note that for BNS merger we only consider the disk wind (since there are polar shocked ejecta, the wind-only treatment may be a lower limit to the real $M_w$ in this case).   

%%%%%%%%%%%%%%%%%%%%%%%%%%%%%%%%%%%%%%%%%%%%%%%%%%%%figure_5

\begin{figure*}[ht]
  \centering
    \subfigure{
    \begin{minipage}[t]{8.6cm}
    \includegraphics[width=8.6cm]{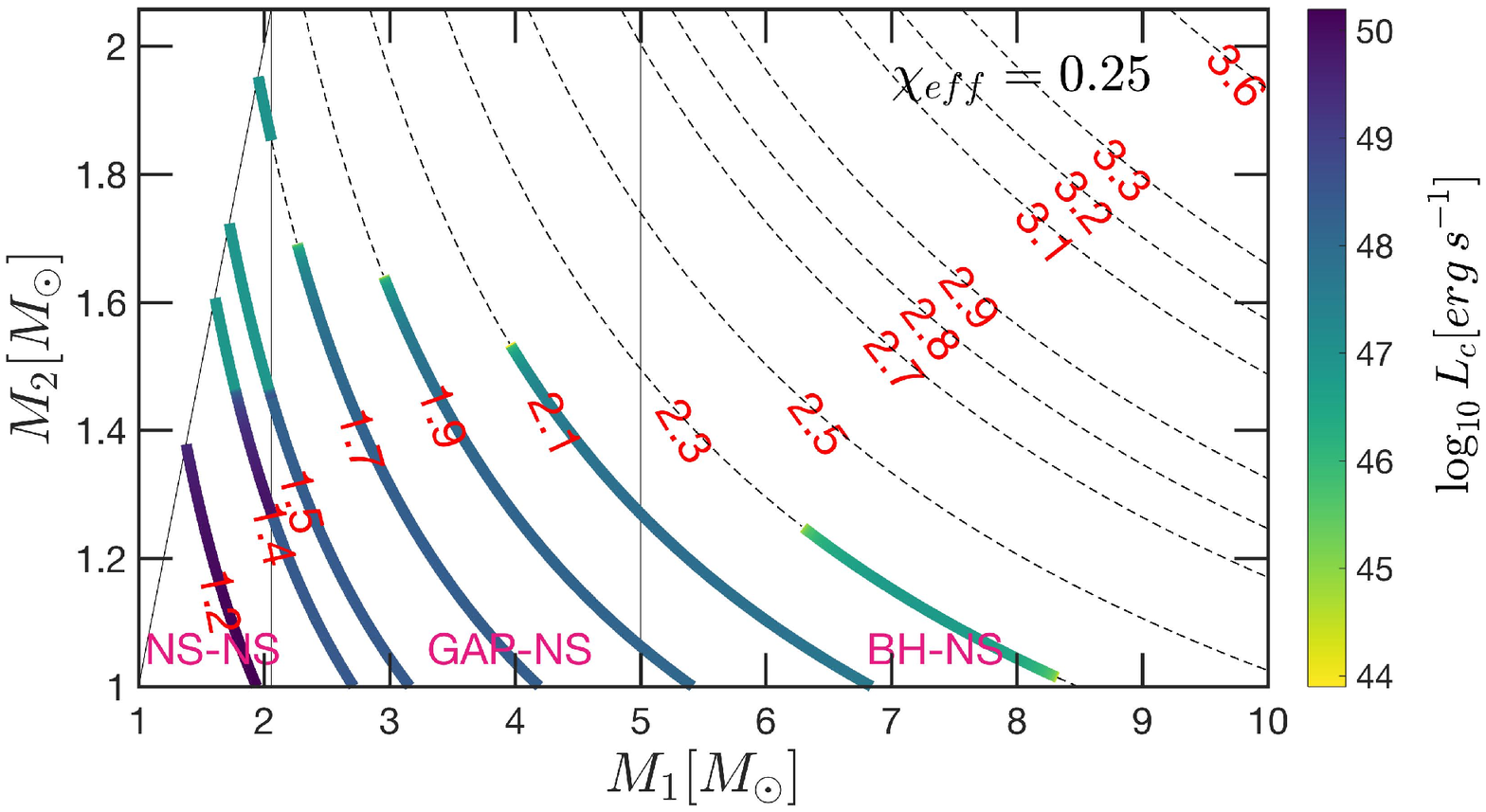}
    \end{minipage}} 
    \subfigure{
    \begin{minipage}[t]{8.6cm}
    \includegraphics[width=8.6cm]{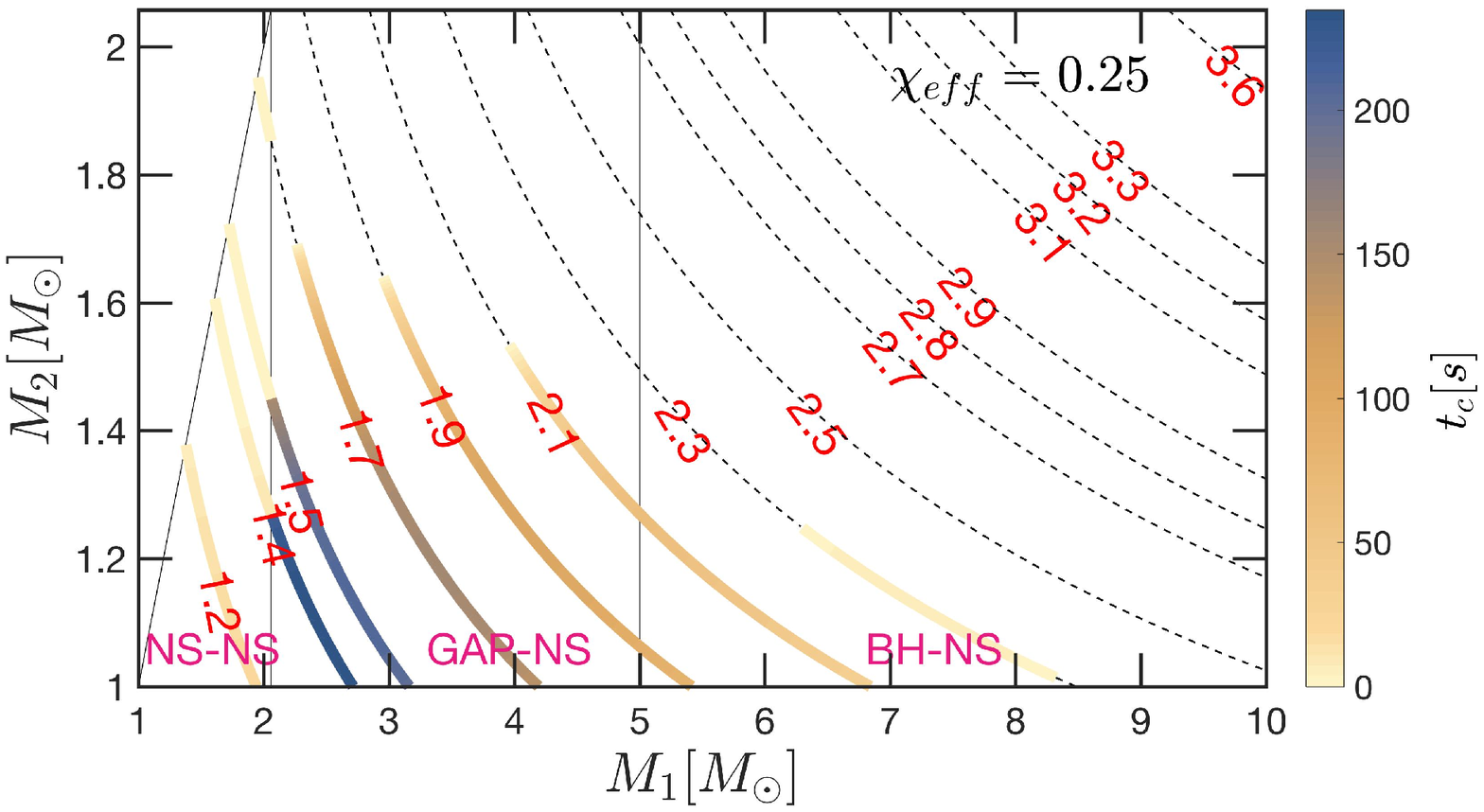}
    \end{minipage}} 
    	
    \subfigure{
    \begin{minipage}[t]{8.6cm}
    \includegraphics[width=8.6cm]{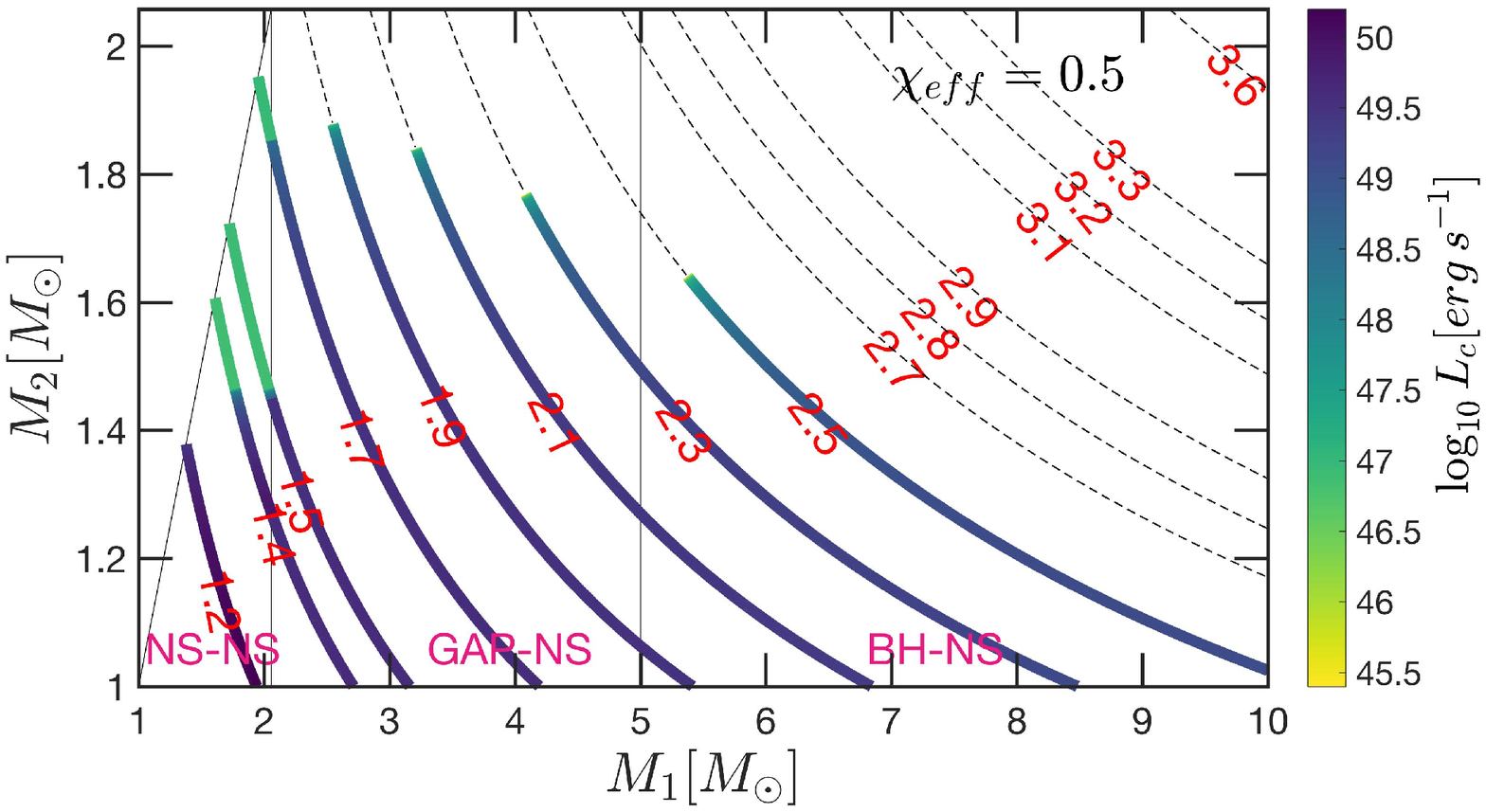}
    \end{minipage}} 
    \subfigure{
    \begin{minipage}[t]{8.6cm}
    \includegraphics[width=8.6cm]{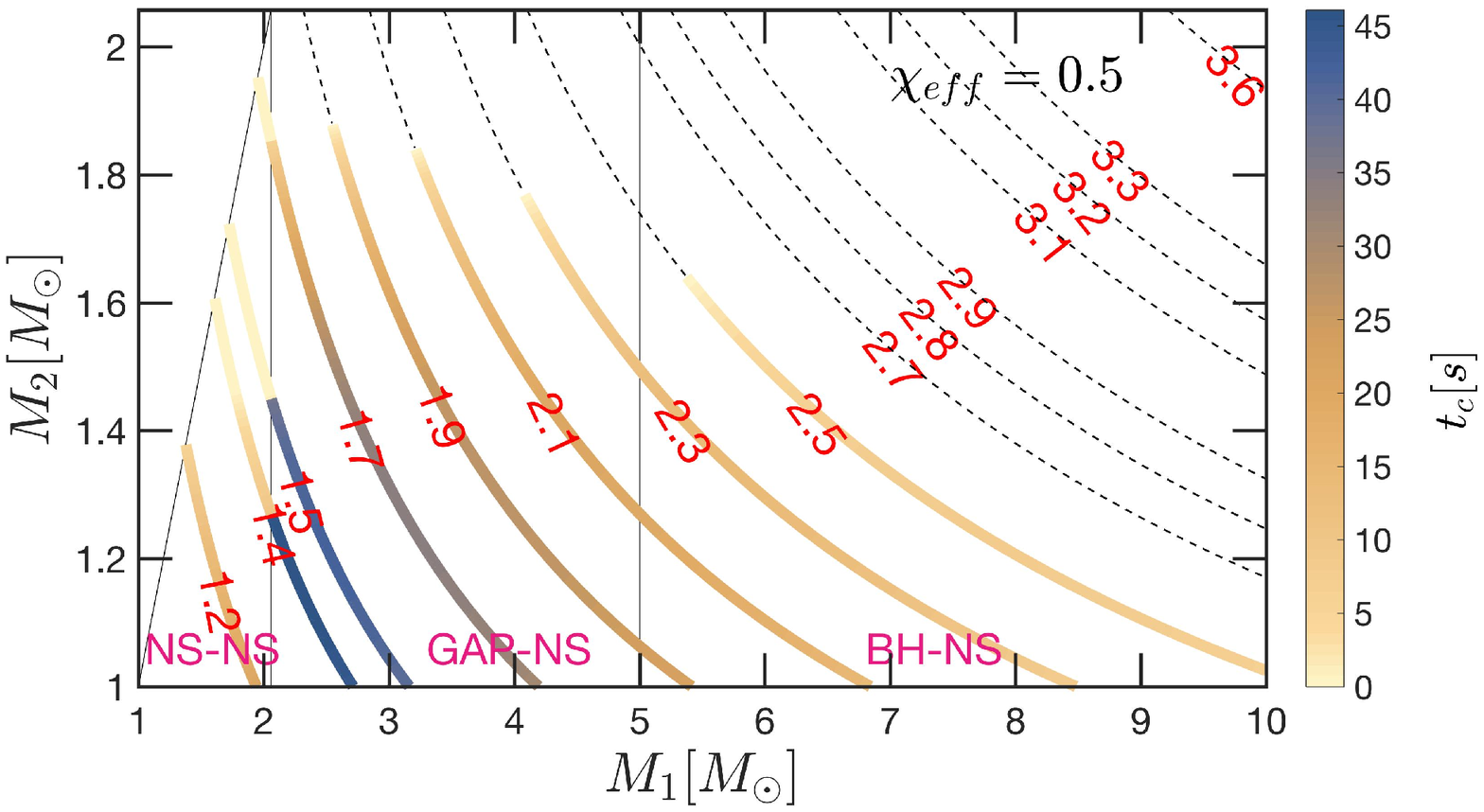}
    \end{minipage}} 
    
    \subfigure{
    \begin{minipage}[t]{8.6cm}
    \includegraphics[width=8.6cm]{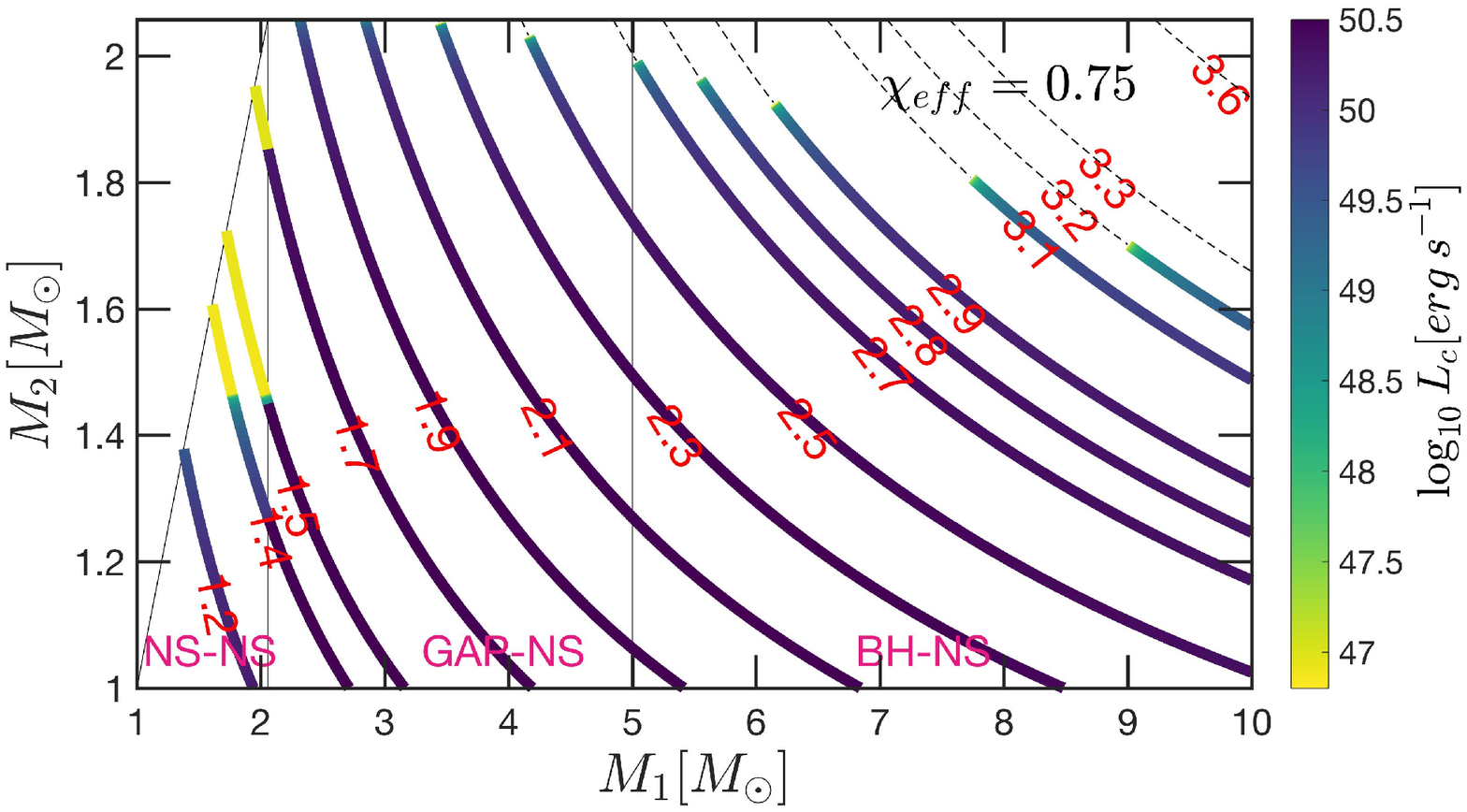}
    \end{minipage}} 
    \subfigure{
     \begin{minipage}[t]{8.6cm}
    \includegraphics[width=8.6cm]{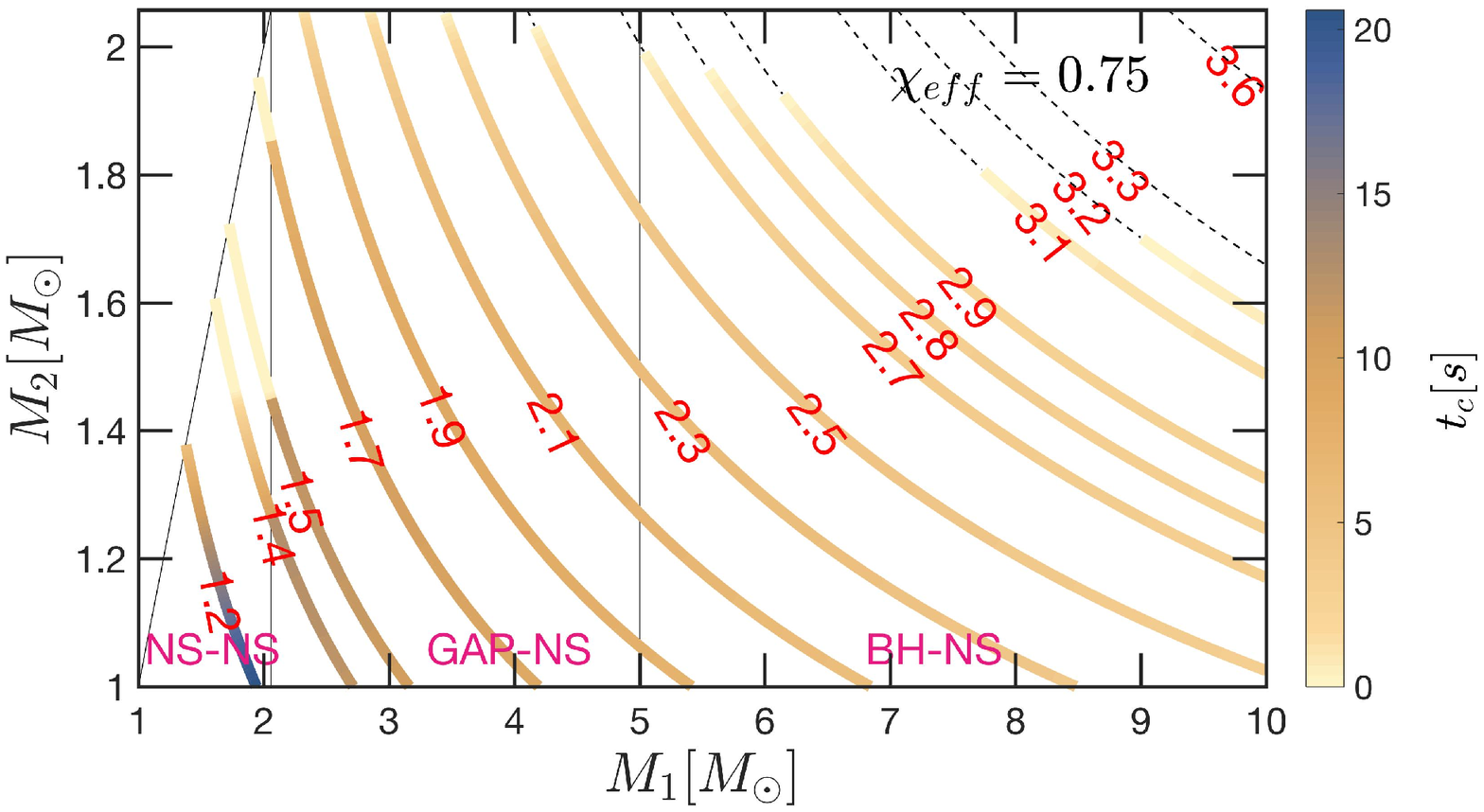}
    \end{minipage}}    
    \centering 
   \caption{Peak luminosity (left) and duration (right) of the cocoon prompt emission for initial effective BH spin of 0.25 (top), 0.5 (middle) and 0.75 (bottom). The pre-burst ejecta is assumed to be disk wind only for all the merging system. The wind ejecta is assumed to be isotropically distributed.}
	\label{fig:cocoon_result}
\end{figure*}

%%%%%%%%%%%%%%%%%%%%%%%%%%%%%%%%%%%%%%%%%%%%%%%%%%%%%figure_5

In the homologously expanding limit, \cite{duffell18} found that jets are successful when ${E_{\rm{j}}} \gtrsim 0.1{E_{{\rm{w}}}}$ where ${E_{{\rm{w}}}} \approx {M_{{\rm{w}}}}\beta _{{\rm{w}}}^2{c^2}/2$ is the kinetic energy of the ejecta and $E_j$ denotes the total energy of the jet. They identified two breakout regimes. For energies in the range $0.1{E_{{\rm{w}}}} \lesssim {E_{\rm{j}}}  \lesssim 3{E_{{\rm{w}}}}$, jets barely break out and a significant amount of energy is deposited in a cocoon. This regime is dubbed the `late breakout'. For higher energies, ${E_{\rm{j}}} \gtrsim 3{E_{{\rm{w}}}}$, jets break out easily, and this regime is dubbed `early breakout'. The jet breakout time is given by \cite{duffell18} as follows
\begin{equation}
{t_{{\rm{b,e}}}} = \left\{ {\begin{array}{*{20}{c}}
{0.3{t_{\rm{e}}}\frac{{{E_{\rm{w}}}}}{{{E_{\rm{j}}}}}{\rm{ = }}0.15{\eta _\gamma }\frac{{{M_{\rm{w}}}{{\left( {{\beta _{\rm{w}}}c} \right)}^2}}}{{{L_{{\rm{\gamma}}}}}}}&{'early'}\\
{\frac{{9{t_{\rm{e}}}}}{{\sqrt {\frac{{10{E_{\rm{j}}}}}{{{E_{\rm{w}}}}} - 1} }} = \frac{{9{t_{\rm{e}}}}}{{\sqrt {\frac{{20{L_{{\rm{\gamma}}}}{t_{\rm{e}}}}}{{{\eta _\gamma }{M_{\rm{w}}}{{\left( {{\beta _{\rm{w}}}c} \right)}^2}}} - 1} }}}&{'late'}
\end{array}} \right. 
\end{equation}
where $t_e\approx{t_{visc}}$ is the duration of the jet engine operation.

\cite{beniamini20} assume that the self-collimation of sGRB jets does not play a critical role and find that the time interval between the binary merger and the launching of a typical sGRB jet is $\leq$ 0.1 s. In this paper, we assume the time $t_w$ is equal to 0.1 s for all the sGRB cases. Then using Eqs. (26)--(30) with $t_w = 0.1s$, we calculate the jet breakout time, $t_{b}$, for the mergers with given chirp masses. 

It is worth noting that the total energy of the cocoon produced by two-sided jets can be approximated as \citep{nakar17} 
\begin{equation}
{E_{\rm{c}}} \approx 2\times{L_{\rm{j}}}\times{t_{\rm{b}}}\end{equation}
and the distance that the jet has to travel in the ambient ejecta can be expressed as \citep{lazzati17}
\begin{equation}
R_a = t_{b} \times \beta_{h,e} \times c
\end{equation}
where $\beta _{{\rm{h,e}}}= \frac{{{\beta _{\rm{w}}} + {\beta _{{\rm{h}}}}}}{{1 + {\beta _{\rm{w}}}{\beta _{{\rm{h}}}}}}$ denotes the jet head velocity in the co-moving frame of the medium \citep{lazzati19}. Thus using the estimated jet breakout time with the method discussed above, one can get the value of $E_c$ and $R_a$ for different merger events.   

%%%%%%%%%%%%%%%%%%%%%%%%%%%%%%%%%%%%%%%%%%%%%%%%%%%%%%%%%%%
\subsection{Lorentz factor of cocoon}

\cite{nakar17} studied the cocoon dynamics following the jet breakout. The terminal Lorentz factor of the cocoon is ${\Gamma _{{\rm{c}}}} = \min \left\{ {{n_{{\rm{c}}}},{\eta _{\rm{b}}}} \right\}$, where $\eta_b$ is the critical baryonic loading, and $\eta_{c}$ is the usual notation of baryonic loading. In our calculation, $\eta_b$ is much larger than $\eta_{c}$ therefore we use
\begin{equation}
\Gamma_c={\eta _{{\rm{c}}}} \equiv \frac{{{E_{{\rm{c}}}}}}{{{m_{{\rm{c}}}}{c^2}}}
\end{equation}
to estimate $\Gamma_c$, where $E_c$ is the energy deposited to the cocoon and $m_{c}$ is its mass. We estimate $m_{c}$ with the relation $m_{c}$ = $\rho_w$ $\times$ $V_c$. The cocoon shape at the time of breakout, as seen in numerical simulations \citep{morsony07, mizuta13} and predicted by analytic modeling \citep{bromberg11}, is roughly a cone or a cylinder with a height $R_a$, thus the volume of cocoon can be estimated as \citep{nakar17}
\begin{equation}
 {V_{\rm{c}}} = \frac{1}{3}\pi R_ a ^3\theta _{\rm{j}}^2 
 \end{equation}
where $\theta_j$ is the half-opening angle of jet. The density of the wind ejecta is approximated as ${\rho _w} = \frac{{{{\dot m}_{\rm{w}}}}}{{{\Omega _{\rm{w}}}R_ a ^2{v_{\rm{w}}}}}$ where $\Omega_w$ is equal to $2\pi$ for the isotropic distribution of wind ejecta \citep{lazzati19}.

\cite{lazzati19} derive analytic estimates for the structure of jets expanding in environments with different density, velocity, and radial extent. They obtain the jet solid angle as
\begin{equation}
{\Omega _{\rm{j}}} = \sqrt {\frac{{3\pi {L_{\rm{j}}}{v_{\rm{w}}}{\Omega _{\rm{w}}}}}{{{{\dot m}_{\rm{w}}}{c^2}{v_{{\rm{h,e}}}}}}} {\sin ^2}{\theta _{{\rm{j,inj}}}}
\end{equation}
where $\theta_{j,inj}$ is the injection angle of jet. Thus with $\Omega_j=2\pi(1-cos\theta_j)$, the value of $\theta_j$ for different mergers can be obtained. Using Eqs. (33) -- (35), we derive the value of $\Gamma_c$ for any given merger.  

%%%%%%%%%%%%%%%%%%%%%%%%%%%%%%%%%%%%%%%%%%%%%%%%%%%%%%%%%%%%%
\subsection{Results of cocoon prompt emission}  

Figure \ref{fig:cocoon_result} shows the variation of peak luminosity $L_c$ (left) and duration $t_c$ (right) of the cocoon prompt emssion versus the parameters of merging system. Note that for all mergers, we set the injection angle of jet to be $10^{\circ}$. 

%%%%%%%%%%%%%%%%%%%%%%%%%%%%%%%%%%%%%%%%%%%%%%%%%%%%figure_6

\begin{figure*}[htb]
  \centering
    \subfigure{
    \begin{minipage}[t]{8.6cm}
    \includegraphics[width=8.6cm]{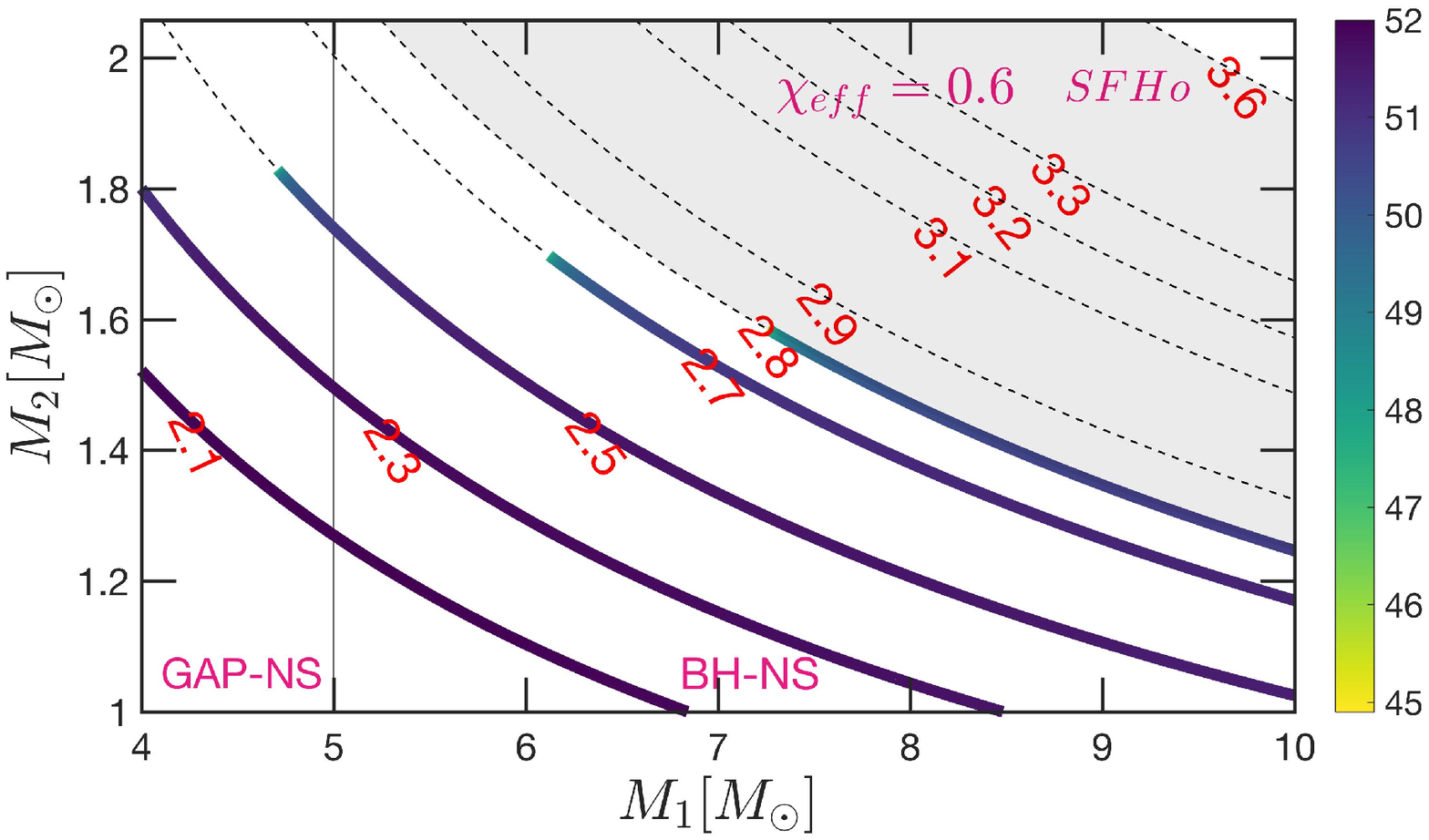}
    \end{minipage}} 
    \subfigure{
    \begin{minipage}[t]{8.6cm}
    \includegraphics[width=8.6cm]{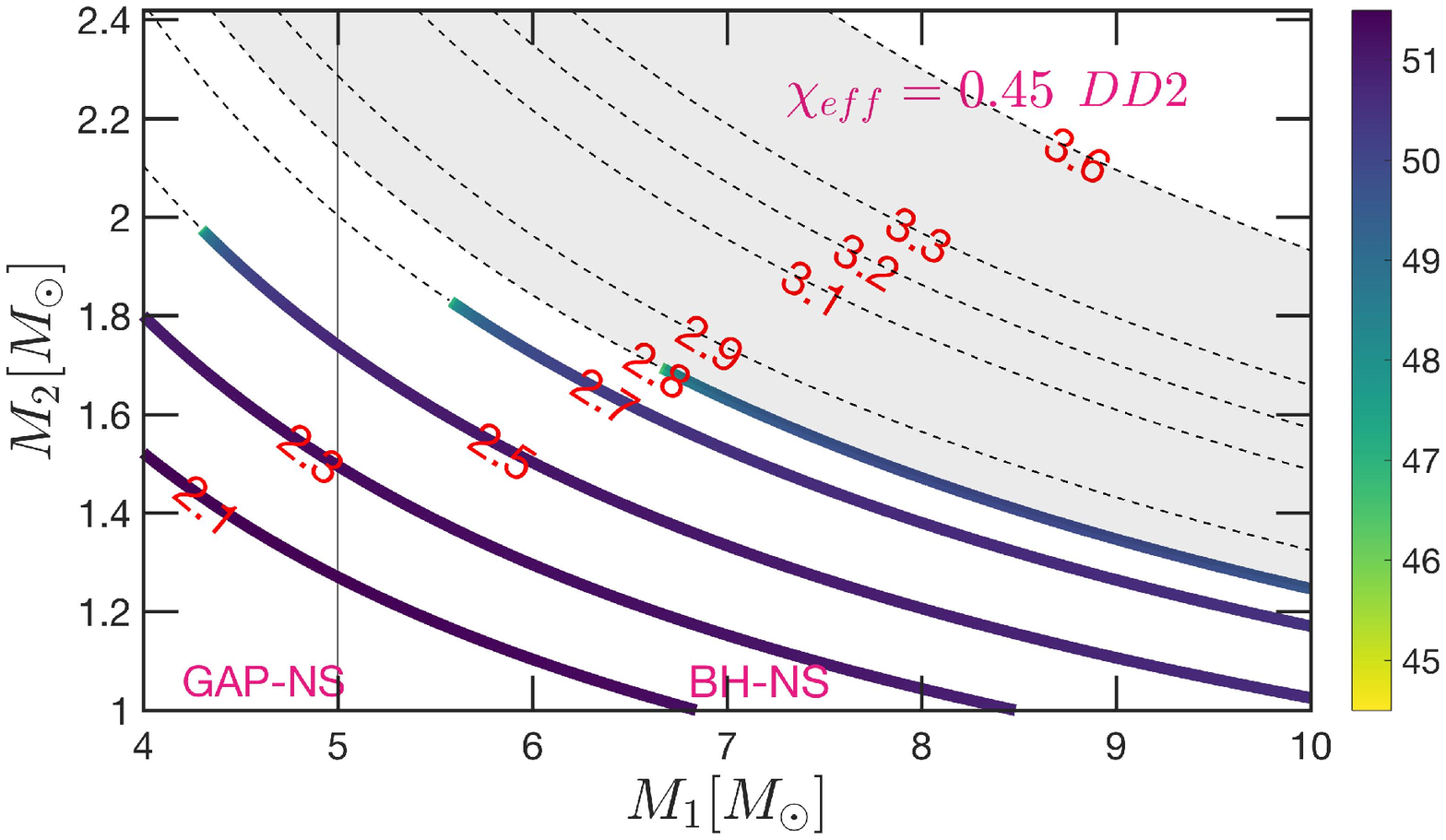}
    \end{minipage}} 
    	
    \subfigure{
    \begin{minipage}[t]{8.6cm}
    \includegraphics[width=8.6cm]{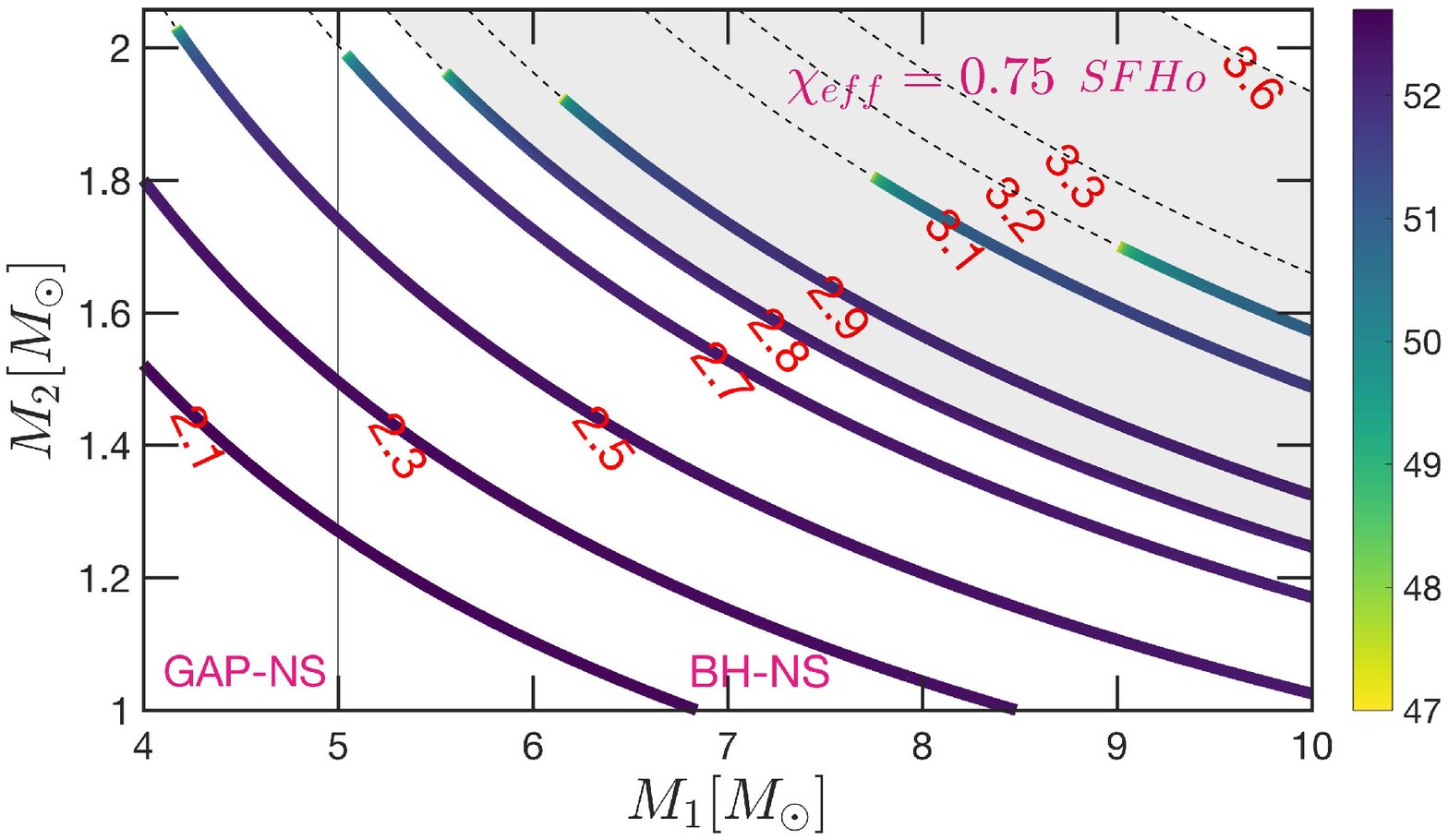}
    \end{minipage}} 
    \subfigure{
    \begin{minipage}[t]{8.6cm}
    \includegraphics[width=8.6cm]{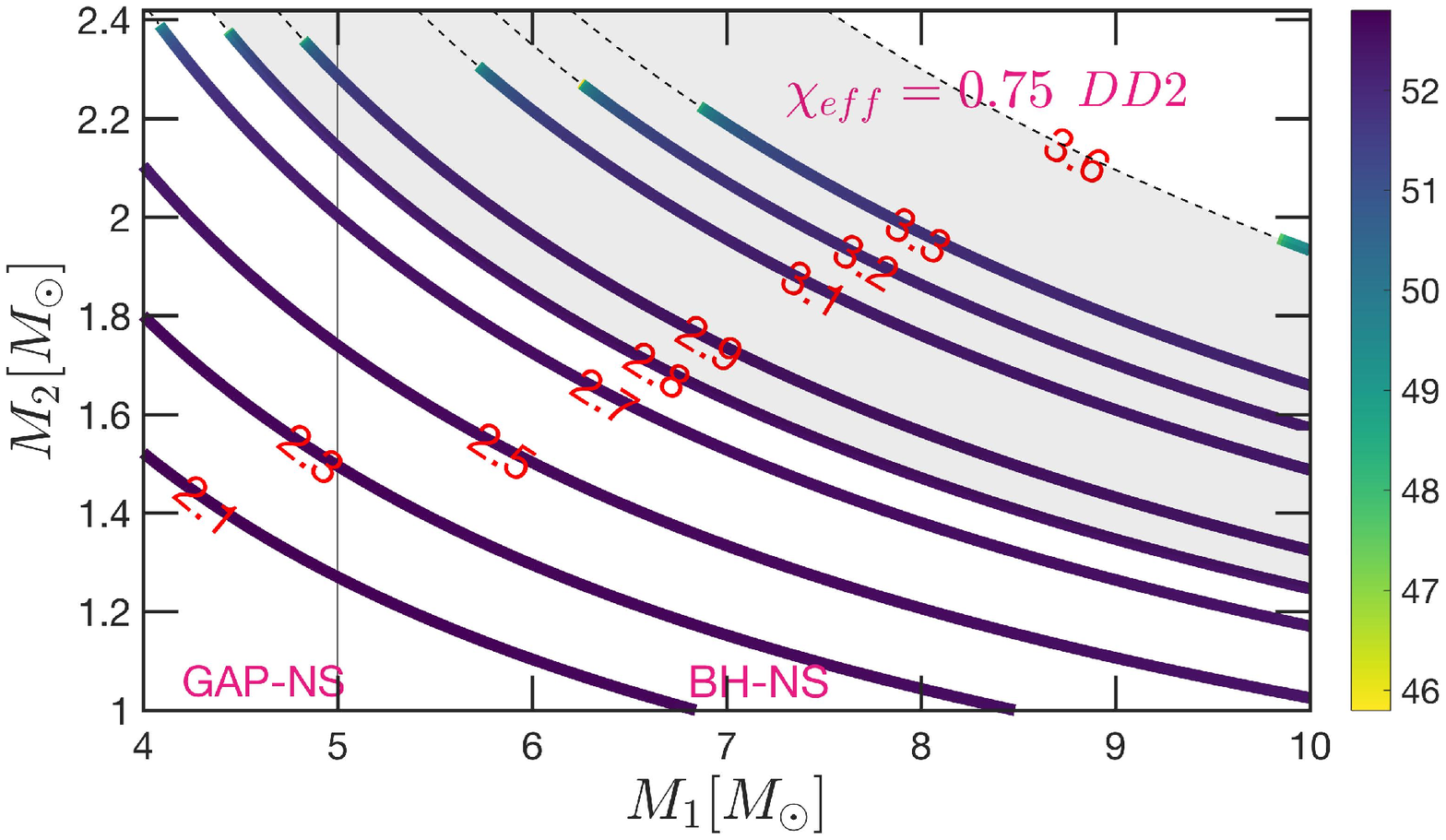}
    \end{minipage}} 
    
    % \subfigure{
    % \begin{minipage}[t]{8.6cm}
    % \includegraphics[width=8.6cm]{{L_COC_spin_0.75}.eps}
    % \end{minipage}} 
    %\subfigure{
    % \begin{minipage}[t]{8.6cm}
    %\includegraphics[width=8.6cm]{{t_COC_spin_0.75}.eps}
    %\end{minipage}}    
    \centering 
   \caption{The peak isotropic equivalent gamma-rays luminosity of sGRB prompt emission for EOS SFHo (left) and DD2 (right) (see the upper-right corner of each panel for the initial spin of the BH). Note that the maximum mass of neutron stars for DD2 EOS is $2.42M_{\odot}$, thus the maximum value of $M_2$ in the right column is higher than that in the left column. The light gray region represents 90\% confidence interval of the chirp mass for GRB GBM--190816 (see section 7).}
	\label{fig:eos_result}
\end{figure*}

%%%%%%%%%%%%%%%%%%%%%%%%%%%%%%%%%%%%%%%%%%%%%%%%%%%%%figure_6

As shown in Figure \ref{fig:cocoon_result}, for the cocoon prompt emission, which peaks in X-ray and UV bands, $L_c$ is mainly in the range of $10^{45}$ to $10^{51}$ erg/s. The merging system with lower effective spin would have a larger $t_c$, as $t_c$ decreases with $\Gamma_c$ which is positively related to $t_{b}$ and $M_{w}$. For a given chirp mass ranging from $1.5M_{\odot}$ to $1.7M_{\odot}$, the peak luminosities of BNS mergers are lower than those of GAP-NS mergers by one to three orders of magnitude. Besides, the duration of cocoon prompt emission from BNS mergers are also lower than those of GAP-NS mergers. These differences would be useful to distinguish the two types of mergers. It is worth noting that in this range of chirp mass, the cocoon prompt emission is likely to be detected only for GAP--NS events because $t_c$ is too short to be detected for other cases according to current detection limit.

%%%%%%%%%%%%%%%%%%%%%%%%%%%%%%%%%%%%%%%%%%%%%%%%%%%%%%%%%%%%%%%%%%%%%%%%%%
\section{Case study for GRB GBM--190816}

Fermi GBM-190816, a sub-threshold GRB candidate, was potentially associated with a sub-threshold LIGO/Virgo compact binary merger candidate, as reported by \cite{lvf19} and \cite{goldstein19}. According to the GW signal, the lighter compact object is estimated to be lighter than 3$M_{\odot}$, which can be either an NS or a low-mass BH. The heavier component is a higher-mass BH. 

\cite{yang20} perform an independent analysis of the publicly available data and investigate the physical implications of this potential association. Taking into account the burst distance $\sim$ 428 Mpc, they calculate the isotropic $\gamma$-ray luminosity as ${L_{{\rm{\gamma }},{\rm{iso}}}}{\rm{ = }}1.47_{ - 1.04}^{ + 3.40} \times {10^{49}}{\rm{erg}}\,{{\rm{s}}^{ - 1}}$. Based on their assumption that one compact object of this CBC event is an NS with a mass of 1.4$M_{\odot}$, using the GW date alone, they further constrain the mass ratio to be $q = 2.26_{ - 0.12}^{ + 2.75}$. 

These component masses give the chirp mass of this event $M_{\rm ch} = 2.9_{ - 0.1}^{ + 0.7}M_{\odot}$ (90\% confidence interval, shown as the light gray region in Figure \ref{fig:eos_result}). Thus combining $L_{\gamma,iso}$ and $M_{ch}$ of GBM-190816 with our estimation scheme above, we find that the effective spin should be larger than 0.6, for the EOS SFHo (see the left column of Figure \ref{fig:eos_result}). For a given $\chi_{eff}$, e.g., $\chi_{eff}=0.75$, the left bottom panel in Figure \ref{fig:eos_result} suggests that the mass of the lighter one should be smaller than 2$M_{\odot}$ and the mass of the BH component should be larger than 5.5$M_{\odot}$ for the EOS SFHo.

As the neutron star EOS is a major source of uncertainty for the ejecta properties, we also take a much stiffer EOS (i.e., DD2, whose maximum mass of neutron star is $M_{{\rm{NS}}}^{\max }=2.42M_{\odot}$) from \cite{hempel10} and \cite{typel10} to approximately capture the edge cases of the parameter space. As a stiffer EOS would result in more ejecta, for the DD2 EOS, the inferred minimum value of $\chi_{eff}$ = 0.45 (see the right column of Figure \ref{fig:eos_result}) is smaller than that inferred from the case with SFHo EOS. Besides, compared with the SFHo EOS, for the given $\chi_{eff}=0.75$, the DD2 EOS results in a larger upper mass limit of the lighter one (2.4$M_{\odot}$, close to the maximum mass of neutron star for the DD2 EOS) and a smaller lower mass limit of the BH component (4.5$M_{\odot}$).

%%%%%%%%%%%%%%%%%%%%%%%%%%%%%%%%%%%%%%%%%%%%%%%%%%%%%%%%%%%%%%%%%%%%%%%%%%%
\section{Summary and Discussion}

The detection of electromagnetic radiation from a compact binary merger triggered by GWs plays role on studying the merging system including the properties of the outflow, the physics of the relativistic jet and the interaction between the jet and the pre-burst ejecta. In this paper we investigate the early EM emission which is critical in recognizing the merger and give a detailed estimation for kilonova, sGRB and cocoon prompt emission. We connect our calculation for those early emission components with the chirp mass $M_{ch}$ which is one of the best measured parameters encoded in the GW signal.

Based on the fitting formulae for the disk and dynamical ejecta masses of BNS and BH--NS mergers from numerical simulations \citep{foucart18, kruger20}, we found that the mass of disk and dynamical ejecta can be up to 0.3$M_{\odot}$ and 0.1$M_{\odot}$ for the SFHo EOS, respectively (Figure \ref{fig:mass_result}). For BH--NS mergers, the amount of two types of ejecta increases with the initial effective spin of the merging system. For a given chirp mass, the disruption of NS tends to occur in the system with lower mass of NS (therefore, larger mass of BH). The mass of disk is larger than that of dynamical ejecta for any given merger.

\cite{barbieri19a} pointed out that the chirp mass range of $1.2M_{\odot}-1.8M_{\odot}$ is compatible to both a BNS and a GAP-NS mergers. In this work, we found that for a given chirp mass ranging from $1.5M_{\odot}$ to $1.7M_{\odot}$, the disk mass of a BNS merger is far lower than that of a GAP-NS merger. Besides, all the BNS mergers in this range of chirp mass hardly eject dynamical ejecta while some of the GAP-NS mergers eject significant amount of dynamical ejecta around 0.02$M_{\odot}$. Due to the dependence of the kilonova, sGRB, cocoon on the properties of the dynamical ejecta and disk wind, we show that those differences in the dynamical ejecta and disk between BNS mergers and GAP-NS mergers for a given chirp mass at the range of $1.5M_{\odot}-1.7M_{\odot}$ could result in significant differences in those emissions between the two types of mergers.

As shown in Figure \ref{fig:KN_result}, for a given $M_{ch}$ in the range of $1.5M_{\odot}-1.7M_{\odot}$, the peak luminosities of the kilonova emission from BNS mergers are nearly tow orders lower than those of GAP--NS mergers. Those kilonova emissions from BNS mergers tend to be bluer and peak earlier than GAP--NS mergers. Therefore, the observation of kilonova (i.e., peak luminosity and timescale) from the mergers whose chirp mass is in the range of about $1.5M_{\odot}-1.7M_{\odot}$ can help identify those mergers. This is consistent with the result of \cite{barbieri19a}, in which they studied the light curves of both BNS and BH--NS mergers and pointed out that at the optimal value of chirp mass ($M_{ch}=1.45M_{\odot}$) a single observation in the g or K band would be helpful to distinguish the nature of the merging system.

For sGRBs powered by accretion onto the BH, the isotropic equivalent $\gamma$-rays luminosity is sensitive to the disk mass. For $M_{ch}=1.5M_{\odot}-1.7M_{\odot}$ the BNS mergers eject less disk wind compared with GAP--NS mergers, thus $L_{\gamma}$ of BNS merger is lower than that of GAP--NS merger by around three to four orders of magnitude (see Figure \ref{fig:LpeakGRB}). Therefore it can be concluded that the observation of sGRB may also be helpful to identify those mergers when the detected chirp mass falls in the range of $1.5M_{\odot}-1.7M_{\odot}$.

As shown in Figure \ref{fig:cocoon_result}, for a given chirp mass ranging from $1.5M_{\odot}$ to $1.7M_{\odot}$, the cocoon prompt emission which peaks in X-ray and UV band form BNS merger is different from that of GAP-NS merger. The peak luminosities of BNS mergers are lower than those of GAP-NS mergers by one to three orders of magnitude (note that this difference increases with spin). Besides, the duration of cocoon prompt emission from BNS mergers are also lower than those of GAP-NS mergers. Therefore, except for the kilonova and sGRB emissions, the cocoon prompt emission can also be used to reveal the nature of a merger.

The early emissions estimated in this work are useful to constrain the properties of merger events (see section 7 for example). When $M_{ch}$ falls in the range of $1.5M_{\odot}-1.7M_{\odot}$ which is compatible to both a BNS and a GAP--NS mergers, we find that the mass of the dynamical ejecta and disk of the BNS merger is far lower than that of the GAP--NS merger, which would result in the differences in kilonova, sGRB and cocoon between BNS mergers and GAP--NS mergers. These are useful in distinguishing the two merger types when the detected chirp mass falls in the range of $1.5M_{\odot}-1.7M_{\odot}$.

%%%%%%%%%%%%%%%%%%%%%%%%%%%%%%%
\section*{}

We thank the anonymous referee for useful comments and suggestions that improve the quality of the paper. This work is supported by National Natural Science Foundation of China (11673078 and 12073091), Guangdong Basic and Applied Basic Research Foundation (2019A1515011119) and Guangdong Major Project of Basic and Applied Basic Research (2019B030302001).
%%%%%%%%%%%%%%%%%%%%%%%%%%%%%%%%%%%%%%%%%%%%%%%%%%%%%%%
%\newpage
%\newpage

%%%%%%%%%%%%%%%%%%%%%%%%%%%%%%
\end{CJK*}

\begin{thebibliography}{99}
\bibitem[Abbott et al.(2017a)]{abbott17a} Abbott, B. P., et al. 2017a, ApJL, 848(2): L12

\bibitem[Abbott et al.(2017b)]{abbott17b} Abbott, B. P., et al., 2017b, ApJL, 848(2): L13

%\bibitem[Abbott et al.(2020a)]{abbott20a} Abbott, B. P., et al., 2020a, ApJL, 896: L44

\bibitem[Abbott et al.(2020)]{abbott20} Abbott, B. P., et al., 2020b, 	arXiv:2010.14533

\bibitem[Arnett(1982)]{arnett82} Arnett, W. D. 1982, ApJ, 253, 785

\bibitem[Bailyn et al.(1998)]{bailyn98} Bailyn, C. D., Jain, R. K., Coppi, P., \& Orosz, J. A. 1998, ApJ, 499, 367

\bibitem[Barbieri et al.(2019a)]{barbieri19a} Barbieri, C., Salafia, O. S., Colpi, M., Ghirlanda, G., Perego, A., Colombo, A. 2019, ApJL, 887(2):L35

\bibitem[Barbieri et al.(2019b)]{barbieri19b} Barbieri, C., Salafia, O. S., Perego, A., Colpi, M., \& Ghirlanda, G. 2019, A\&A, 625, A152

\bibitem[Barnes \& Kasen(2013)]{barnes13} Barnes, J., \& Kasen, D. 2013, ApJ, 775(1):18  

\bibitem[Barnes et al.(2016)]{barnes16} Barnes, J., Kasen, D., Wu, Meng-Ru., Mart\'{i}nez-Pinedo, Gabriel. 2016, 829(2):110 

\bibitem[Bauswein et al.(2013)]{bauswein13} Bauswein, A., Goriely, S., Janka, H.T. 2013, ApJ, 773:78

\bibitem[Bavera et al.(2019)]{bavera19} Bavera, S. S., Fragos, T., Qin, Y., et al. 2019, A\&A, 635, A97

\bibitem[Begelman \& Cioffi(1989)]{begelman89} Begelman, M. C., \& Cioffi, D. F. 1989, ApJL, 345, L21

\bibitem[Beniamini et al.(2020)]{beniamini20} Beniamini, Paz; Duran, Rodolfo Barniol; Petropoulou, Maria; Giannios, Dimitrios. 2020, The ApJL, 895(2):L33  

\bibitem[Bhattacharya, Kumar \& Smoot(2019)]{bhattacharya19} Bhattacharya, M., Kumar, P., \& Smoot, G. 2019, MNRAS, 486, 5289  

\bibitem[Blandford \& Znajek(1977)]{blandford77} Blandford, R. D., \& Znajek, R. L., 1977, MNRAS, 179, 433-456 

\bibitem[Bloom et al.(2009)]{bloom09} Bloom, J. S., Holz, D. E., Hughes, S. A., et al. 2009, e-prints arXiv:0902.1527

\bibitem[Bovard et al.(2017)]{bovard17} Bovard, Luke;  Martin, Dirk;  Guercilena, Federico; et al. 2017,  PhRvD, 96,12, 124005

\bibitem[Bromberg et al.(2011)]{bromberg11} Bromberg, O., Nakar, E., Piran, T., \& Sari, R. 2011, ApJ, 740, 100

%\bibitem[Daniel(1980)]{daniel80} Daniel, J. Y., 1980, A\&A, 86, 198  

\bibitem[Coughlin et al.(2019)]{coughlin19} Coughlin M. W.; Dietrich T.; Margalit B.; Metzger B. D., 2019, MNRAS, 489, L91

\bibitem[Dietrich \& Ujevic(2017)]{dietrich17} Dietrich, Tim; Ujevic, Maximiliano, 2017, Classical and Quantum Gravity, 34, 10, 105014 

\bibitem[Doctor et al.(2019)]{doctor19}Doctor, Z., Wysocki, D., O’Shaughnessy, R., Holz, D. E., \& Farr, B. 2019, ApJ, 893, 35

\bibitem[Duffell et al.(2018)]{duffell18} Paul C. Duffell; Eliot Quataert; Daniel Kasen; Hannah Klion, 2018, ApJ, 866:3

\bibitem[Eichler et al.(1989)]{eichler89} Eichler D., Livio M., Piran T., Schramm DN. 1989, Nature 340:126–128.

\bibitem[Farr et al.(2011)]{farr11} Farr, Will M.;  Sravan, Niharika;  Cantrell, Andrew; et al. 2011, ApJ, 741, 103

\bibitem[Fern\'{a}ndez \& Metzger(2013)]{fernandez13} Fern\'{a}ndez, R., \& Metzger, B. D., 2013, MNRAS, 435, 1, 502-517

\bibitem[Fern\'{a}ndez et al.(2015a)]{fernandez15a} Fern\'{a}ndez, Rodrigo;  Quataert, Eliot;  Schwab, Josiah;  Kasen, Daniel;  Rosswog, Stephan. 2015, MNRAS, 449, 1, 390-402

\bibitem[Fern\'{a}ndez et al.(2015b)]{fernandez15b} Fern\'{a}ndez, Rodrigo;  Kasen, Daniel;  Metzger, Brian D.;  Quataert, Eliot, 2015, MNRAS 446, 750–758

\bibitem[Fern\'{a}ndez \& Metzger(2016)]{fernandez16} Fern\'{a}ndez, R.; Metzger, B. D., 2016, Annu. Rev. Nucl. Part. Sci., 66, 23

\bibitem[Fern\'{a}ndez et al.(2017)]{fernandez17} Fern\'{a}ndez, R., Foucart, F., Kasen, D., et al. 2017, Classical and Quantum Gravity, 34, 15, 154001 

\bibitem[Fern\'{a}ndez et al.(2019)]{fernandez19} Fern\'{a}ndez, R., Tchekhovskoy, A., Quataert, E., Foucart, F., Kasen, D. 2018, MNRAS, 482, 3373-3393.

\bibitem[Fong et al.(2014)]{fong14} Fong W., Berger E., Metzger BD., et al. 2014, ApJ, 780:118.

\bibitem[Foucart(2012)]{foucart12} Foucart, Francois, 2012, PhRvD, 86, 12, 124007

\bibitem[Foucart et al.(2014)]{foucart14} Foucart, F., Deaton, M. B., Duez, M. D., et al. 2014, PhRvD, 90, 2, 024026

\bibitem[Foucart et al.(2018)]{foucart18} Foucart, F., Hinderer, T., \& Nissanke, S. 2018, PhRvD, 98, 8, 081501

\bibitem[Frail et al.(2001)]{frail01} Frail, D. A., Kulkarni, S. R., Sari, R. Djorgovski, et al. 2001, ApJ, 562, 1, L55-L58.

\bibitem[Gerosa rt al.(2018)]{gerosa18}Gerosa, D., Berti, E., O’Shaughnessy, R., et al. 2018, PhRvD, 98, 084036

\bibitem[Goldstein et al.(2019)]{goldstein19} Goldstein, A., Hamburg, R., Wood, J., et al. 2019, arXiv e-prints, arXiv:1903.12597

\bibitem[Grossman et al.(2014)]{grossman14} Grossman, D., Korobkin, O., Rosswog, S., Piran, T., 2014, MNRAS, 439, 1, 757-770 

\bibitem[Hempel \& Schaffner-Bielich(2010)]{hempel10} Hempel, M., \& Schaffner-Bielich, J. 2010, NuPhA, 837, 210

\bibitem[Hotokezaka et al.(2013b)]{hotokezaka13b} Hotokezaka, K., Kiuchi, K., Kyutoku, K., et al. 2013, PhRvD, 88, 4, 044026

\bibitem[Hotokezaka et al.(2013a)]{hotokezaka13a} Hotokezaka, K., Kiuchi, K., Kyutoku, K., Okawa, H., Sekiguchi, Y.i., Shibata, M., Taniguchi, K., 2013, PhRvD, 87, 024001

%\bibitem[Hotokezaka et al.(2018)]{hotokezaka18} Hotokezaka, K.; Kiuchi, K.; Shibata, M.; Nakar, E.; Piran, T., 2018, ApJ, 867(2):95

%\bibitem[Ishii et al.(2018)]{ishii18} Ishii, A., Shigeyama, T., \& Tanaka M. 2018, ApJ, 861, 25

\bibitem[Just et al.(2015)]{just15} Just, O., et al., 2015, MNRAS, 448, 541

%\bibitem[Karp et al.(1977)]{karp77} Karp, A. H., Lasher, G., Chan, K. L., Salpeter, E. E. 1977, ApJ, 214, 161

%\bibitem[Kasen et al.(2003)]{kasen03} Kasen, D., Nugent, P., Wang, L., et al., 2003, ApJ, 593, 788

\bibitem[Kalogera(2000)]{kalogera00}Kalogera, V. 2000, ApJ, 541, 319

\bibitem[Kasen et al.(2013)]{kasen13} Kasen, D., Badnell, N. R., Barnes, J., 2013, ApJ, 774, 25

\bibitem[Kasen et al.(2017)]{kasen17} Kasen, D., Metzger, B.D., Barnes, J., Quataert, E., Ramirez-Ruiz, E., 2017, Nature, 551, 80–84 

%\bibitem[Kasen, Fern\'andez \& Metzger(2015)]{kasen15} Kasen, D., Fern\'andez, R., \& Metzger, B. D., 2015, MNRAS 450, 1777   

\bibitem[Kawaguchi et al.(2015)]{Kawaguchi15} Kawaguchi, K., Kyutoku, K., Nakano, H., Okawa, H., Shibata, M., Taniguchi, K., 2015, PhRvD, 92, 024014 

\bibitem[Kawaguchi et al.(2016)]{kawaguchi16} Kawaguchi, K.,  Kyutoku, K., Shibata, M., Tanaka, M. 2016, ApJ, 825, 1, 52 

\bibitem[Kr\"{u}ger \& Foucart (2020)]{kruger20} Kr\"{u}ger \& Foucart, 2020, PhRvD, 101, 10, 103002

\bibitem[Kiuchi et al.(2015)]{kiuchi15} Kiuchi, K., Sekiguchi, Y., Kyutoku, K., Shibata, M., Taniguchi, K., Wada, T. 2015, PhRvD, 92, 064034

\bibitem[Korobkin et al. (2012)]{korobkin12} Korobkin, O., Rosswog, S., Arcones, A., Winteler, C. 2012, MNRAS, 426, 3, 1940-1949.

%\bibitem[Kulkarni(2005)]{kulkarni05} Kulkarni, S. R., 2005, ArXiv Astrophysics e-prints

\bibitem[Kyutoku et al.(2013)]{kyutoku13} Kyutoku, K., Ioka, K., Shibata, M. 2013, PhRvD, 88, 4, 041503

%\bibitem[Kyutoku et al.(2014)]{kyutoku14} Kyutoku, K., Ioka, K. Shibata, M. 2014, MNRAS, 437, L6

\bibitem[Kyutoku et al.(2015)]{kyutoku15} Kyutoku, K.; Ioka, K.; Okawa, H.; Shibata, M.; Taniguchi, K., 2015, PhRvD, 92, 044028

\bibitem[Kyutoku et al.(2018)]{kyutoku18} Kyutoku, K., Kiuchi, K., Sekiguchi, Y., Shibata, M., Taniguchi, K. 2018, PhRvD, 97, 2, 023009

\bibitem[Lazzati et al.(2017)]{lazzati17} Lazzati, D., Deich, A., Morsony, B. J., Workman, J. C. 2017, MNRAS, 471, 2, 1652-1661

\bibitem[Lazzati et al.(2019)]{lazzati19} Lazzati, D., Perna, R. 2019, ApJ, 881, 2, 89 

\bibitem[Lee et al.(2000)]{lee00} Lee, H. K., Wijers, R. A. M. J., \& Brown, G. E. 2000, PhR, 325, 83

\bibitem[Lei et al.(2005)]{lei05} Lei, W. H., Wang, D. X., \& Ma, R. Y. 2005, ApJ, 619, 420

\bibitem[Lei et al.(2013)]{lei13} Lei, W. H., Zhang, B., \& Liang, E. W. 2013, ApJ, 756, 125 (Paper I)

\bibitem[Lei \& Zhang(2011)]{lei11} Lei, W. H., \& Zhang, B. 2011, ApJL, 740, L27

\bibitem[Li(2000)]{li00} Li, L. X. 2000, PhRvD, 61, 084016

\bibitem[Li \& Paczy\'{n}ski(1998)]{li98} Li, L. X.; \&Paczy\'{n}ski, B.,1998, ApJL, 507:L59-L62.

\bibitem[LIGO/Virgo/Fermi Collaboration.(2019)]{lvf19} LIGO/Virgo/Fermi Collaboration. 2019, GCN, 25406.

\bibitem[Mandel \& O'Shaughnessy (2010)]{mandel10} Mandel, I., \& O'Shaughnessy, R. 2010, CQGra, 27, 114007

\bibitem[Marti et al.(1994)]{marti94} Marti, J. M., Mueller, E., \& Ibanez, J. M. 1994, A\&A, 281, L9

%\bibitem[Matsumoto(2018)]{matsumoto18} Matsumoto, T., 2018, MNRAS, 481, 1, 1008-1015

\bibitem[McKinney(2005)]{mckinney05} McKinney, J. C. 2005, ApJL, 630, L5

\bibitem[Mendoza-Temis et al.(2015)]{mendoza15} Mendoza-Temis, Joel de Jes\'us and Wu, Meng-Ru and Langanke, Karlheinz and Mart\'{\i}nez-Pinedo, Gabriel and Bauswein, Andreas and Janka, Hans-Thomas. 2015, Phys. Rev. C 92, 055805

\bibitem[Metzger et al.(2015)]{metzger15} Metzger, B. D., Bauswein, A., Goriely, S., Kasen, D. 2015, MNRAS, 446, 1115
 
\bibitem[Metzger(2017)]{metzger17} Metzger, B. D., 2017, Living Reviews in Relativity, 20, 3

\bibitem[Matzner(2003)]{matzner03} Matzner, C. D. 2003, MNRAS, 345, 575

\bibitem[Mizuta \& Ioka(2013)]{mizuta13} Mizuta, A., \& Ioka, K. 2013, ApJ, 777, 162

\bibitem[Moderski et al.(1997)]{moderski97} Moderski, R., Sikora, M., Lasota, J. P., 1997, Relativistic Jets in AGNs, Proceedings of the International Conference, p.110-116

\bibitem[Morsony et al.(2007)]{morsony07} Morsony, B. J., Lazzati, D., \& Begelman, M. C. 2007, ApJ, 665, 569

\bibitem[Murguia-Berthier et al.(2014)]{murguia14} Murguia-Berthier, A., Montes, G., Ramirez-Ruiz, E., De Colle, F., Lee, W. H. 2014, ApJL, 788, 1, L8

\bibitem[Murguia-Berthier et al.(2017)]{murguia17} Murguia-Berthier, A., Ramirez-Ruiz, E., Montes, et al. 2017, ApJL, 835(2): L34

\bibitem[Murguia-Berthier et al.(2020)]{murguia20} Murguia-Berthier, A., Ramirez-Ruiz, E., De Colle, F., Janiuk, A. et al. 2020, arXiv:2007.12245

\bibitem[Nagakura et al.(2014)]{nagakura14} Nagakura, H., Hotokezaka, K., Sekiguchi, Y., Shibata, M., Ioka, K. 2014, ApJ, 784, L28

\bibitem[Nakar \& Piran(2017)]{nakar17}Nakar, E., \& Piran, T., 2017, ApJ, 834(1): 28

\bibitem[Narayan et al.(1992)]{narayan92}Narayan R., Paczynski B., \&Piran T. 1992,  ApJL, 395:L83–L86.

\bibitem[Oechslin \& Janka(2006)]{oechslin06} Oechslin, R., Janka, H. T. 2006, MNRAS, 368:1489–1499

\bibitem[Oechslin et al. (2007)]{oechslin07} Oechslin, R., Janka, H. T., Marek, A. 2007, A\&A, 467, 2, 395-409

\bibitem[O'Shaughnessy et al.(2017)]{oshaughnessy17} O'Shaughnessy, R., Gerosa, D., \& Wysocki, D. 2017, PhRvL,119, 011101

\bibitem[\"{O}zel et al.(2010)]{ozel10}\"{O}zel, F., Psaltis, D., Narayan, R., \& McClintock, J. E. 2010, ApJ, 725,1918

\bibitem[Paschalidis(2017)]{paschalidis17} Paschalidis, V., 2017, Classical and Quantum Gravity, 34, 8, 084002

\bibitem[Radice et al.(2016)]{radice16} Radice, D., Galeazzi, F., Lippuner, J., et al. 2016, MNRAS, 460, 3, 3255-3271

\bibitem[Radice et al.(2018)]{radice18} Radice, D., Perego, A., Hotokezaka, K., et al., 2018, ApJ 869:130 

%\bibitem[Radice et al.(2018b)]{radice18b} Radice, D.; Perego, A.; Hotokezaka, K., et al., 2018, arXiv:1809.11163  

\bibitem[Rosswog et al.(1999)]{rosswog99} Rosswog, S., Liebend\"{o}rfer, M., Thielemann, F. K., et al. 1999, A\&A, 341, 499–526

\bibitem[Rosswog et al.(2014)]{rosswog14}Rosswog, S., Korobkin, O., Arcones, A., Thielemann, F. K., Piran, T., 2014, MNRAS, 439:744–756

\bibitem[Ruiz et al.(2018)]{ruiz18} Ruiz, M., Shapiro, S. L., Tsokaros, A. 2018, PhRvD, 97, 2, 021501

\bibitem[Sekiguchi et al.(2016)]{sekiguchi16} Sekiguchi, Y., Kiuchi, K., Kyutoku, K., Shibata, M., Taniguchi, K. 2016, PhRvD, 93, 12, 124046

\bibitem[Shakura \& Sunyaev(1973)]{shakura73} Shakura N. I., Sunyaev R. A., 1973, A\&A, 24, 337

\bibitem[Shapiro(2017)]{shapiro17} Shapiro, S. L., 2017, PhRvD, 95, 10, 101303

\bibitem[Shibata \& Taniguchi(2011)]{shibata11}Shibata, M., \& Taniguchi, K. 2011, Living Rev. Relativity, 14, 6 

\bibitem[Shibata \& Hotokezaka(2019)]{shibata19}Shibata, M., \& Hotokezaka, K., 2019, Annual Review of Nuclear and Particle Science, 69, p.41-64

%\bibitem[Siegel \& Metzger(2017)]{siegel17} Siegel, D. M., Metzger, B. D. 2017, PRL., 119, 1102

%\bibitem[Siegel \& Metzger(2018)]{siegel18} Siegel, D. M., Metzger, B. D. 2018, ApJ, 858, 52

\bibitem[Steiner et al.(2013)]{steiner13}Steiner, A. W., Hempel, M., \& Fischer, T. 2013, ApJ, 774, 17

\bibitem[Tanaka \& Hotokezaka(2013)]{tanaka13} Tanaka, M. \& Hotokezaka, K., 2013, ApJ. 775, 113  

%\bibitem[Tanaka et al.(2014)]{tanaka14} Tanaka, M., Hotokezaka, K., Kyutoku, K., Wanajo, S., Kiuchi, K., Sekiguchi, Y., Shibata, M. 2014, ApJ, 780, 31

\bibitem[Tchekhovskoy et al.(2011)]{tchekhovskoy11} Tchekhovskoy, A., Narayan, R., McKinney, J. C. 2011, MNRAS: Letters, 418, 1, pp. L79-L83.

\bibitem[Thompson et al.(2019)]{thompson19} Thompson, T. A., Kochanek, C. S., Stanek, K. Z., et al. 2019, Sci, 366, 637

\bibitem[Typel et al.(2010)]{typel10} Typel, S., Ropke, G., Klahn, T., Blaschke, D., \& Wolter, H. H. 2010, PhRvC, 81, 015803

\bibitem[Wang et al.(2002)]{wang02} Wang, D. X., Xiao, K., \& Lei, W. H. 2002, MNRAS, 335, 655

%\bibitem[Wanajo et al.(2014)]{wanajo14} Wanajo, S., Sekiguchi, Y., Nishimura, N., Kiuchi, K., Kyutoku, K., Shibata, M. 2014, ApJ, 789, L39

\bibitem[Yang et al.(2020)]{yang20} Yang, Yi-Si., Zhong, Shu-Qing., Zhang, Bin-Bin., et al. 2020, ApJ, 899, 1, 60

\bibitem[Zhang et al.(2018)]{zhang18} Zhang, B-B., Zhang, B., Sun, H., Lei, W-H., et al. 2018, Nat Commun 9, 447

\end{thebibliography}
\end{document}